\documentclass[twocolumn,showpacs,preprintnumbers,prd,floatfix]{revtex4}
\usepackage{epsfig}
\usepackage{array}
\usepackage{graphicx}
\usepackage{dcolumn}% Align table columns on decimal point
\usepackage{bm}
\usepackage{amssymb,amsmath}
\usepackage{natbib}
\usepackage{url}
\usepackage{rotating}
\usepackage{setspace}
\begin{document}
\title{Spinning down newborn neutron stars: nonlinear development of the r-mode instability}
\author{Ruxandra Bondarescu$^1$}
\author{Saul A.\ Teukolsky}
\author{Ira Wasserman}
\affiliation{Department of Physics, Cornell University, Ithaca, New York 14853}
\affiliation{Center for Radiophysics and Space Research, Cornell University, Ithaca, NY 14853, USA}
\begin{abstract}
We model the nonlinear saturation of the r-mode instability via three-mode couplings and the effects 
of the instability on the spin evolution of young neutron stars.  We include one mode triplet 
consisting of the r-mode and two near resonant inertial modes that couple to it. We find that the spectrum of
evolutions is more diverse than previously thought. 
We start our evolutions with a star of temperature $\sim 10^{10}$ K and a spin frequency close to 
the Kepler break-up frequency. We assume that hyperon bulk viscosity dominates at high temperatures (T $\sim$ $10^9-10^{10}$ K) 
and boundary layer viscosity dominates at lower temperatures ($\sim$ a few $\times$ $10^8$ K). To explore possible nonlinear behavior, we vary properties of the star such as the hyperon superfluid transition temperature, the strength of the boundary layer viscosity, and the fraction of the star that cools via direct URCA reactions. The evolution of the star is dynamic and initially dominated by fast neutrino cooling.  Nonlinear effects become important when the r-mode amplitude grows above its first parametric instability threshold. The balance between neutrino cooling and viscous heating plays an important role in the evolution. Depending on the initial r-mode amplitude, and on the strength of the viscosity and of the cooling this balance can occur at different temperatures. If thermal equilibrium occurs on the r-mode stability curve, where gravitational driving equals viscous damping, the evolution may be adequately described by a one-mode model. Otherwise, nonlinear effects are important and lead to various more complicated scenarios. Once thermal balance occurs, the star spins-down oscillating between thermal equilibrium states until the instability is no longer active. The average evolution of the mode amplitudes can be approximated by quasi-stationary states that are determined algebraically. For lower viscosity we observe runaway behavior in which the r-mode amplitude passes several parametric instability thresholds. In this case more modes need to be included to model the evolution accurately. In the most optimistic case, we find that gravitational radiation from the r-mode instability in a very young, fast spinning neutron star within about 1 Mpc of Earth may
be detectable by advanced LIGO for years, and perhaps decades,
after formation. Details regarding the amplitude and duration of the emission depend on the internal dissipation of the modes of the star, which would be probed by such detections.
%We estimate that gravitational radiation from fast spinning young neutron may be detectable by advanced LIGO if such sources exist within 100 kpc.
\end{abstract}
\maketitle
\section{Introduction}
\footnotetext[1]{Current affiliation: Center for Gravitational Wave Physics, Department of Physics, Pennsylvania State University, University Park, PA 16802, USA}
%There are several hundred known core collapse supernovae and more a being discovered. 
%In spite of much progress, 3D core-collapse simulations are complicated and the physics of core-collapse supernova is still not well understood. The common belief that the asymmetry is small and not
%a detectable gravitational wave signal.
Neutron stars are believed to be born in the aftermath of core-collapse supernova explosions as the stellar remnant becomes gravitationally decoupled from the stellar ejecta. Two interesting and timely question are: (1) Do neutron stars spin at millisecond periods at birth or do they spin closer to the observed periods of young pulsars?; (2) Do they emit gravitational radiation that is detectable by
interferometers on Earth? %This year the Laser Interferometer Gravitational-wave Observatory (LIGO) 
%placed upper limits on the gravitational wave emission from the Crab pulsar \cite{Crab} and more accurate gravitational wave searches will target other faster pulsars after the detector is upgraded.

%he collapse of an $8 - 30 M_\odot$ progenitor to a
Theoretically, conservation of angular momentum in the core collapse of a $8 - 30 M_\odot$ progenitor can lead to a newborn neutron star with a period of $\sim 1$ ms or shorter. Observationally, the fastest known young pulsar is in the Large Magellanic Cloud supernova remnant N157B, and has a rotation period of 16 ms \cite{16msDiscovery}. The Crab pulsar is the next fastest neutron star in a supernova remnant with an age $\sim 10^3$ yr.  Its current  rotation period is 33 ms. Its initial period is estimated to be $\sim 19$ ms \cite{WangLai05} by assuming the rotational spin-down is well described by a power law $\dot{\Omega} \propto - \Omega^n$ with braking index $n = 2.51 \pm 0.01$ \cite{lyne}. 
One way to predict the distribution of initial pulsar periods is through population synthesis studies. These studies generally use present day observations with some assumption of their time evolution to reconstruct the birth distribution periods and magnetic fields of the pulsar population. Current studies favor initial periods in the range of several tens to several hundreds of milliseconds \cite{kaspi2006, perna2007}. The apparent discrepancy between the theoretically possible fast rotation rates and the observed slow rotation rates of young neutron stars could be reconciled if the r-mode instability or some other mechanism could spin neutron stars down efficiently, preventing them from maintaining millisecond periods.

 %\footnotetext[1]{An alternative explanation for the relatively low observed spin frequencies of young neutron stars assumes that  the core and the envelope of the progenitor rotate at different rates. If the core loses angular momentum via some mechanism, then the neutron star can be born with a period in the observed range.}

 R-modes are quasi-toroidal oscillations in rotating fluids that occur because of the Coriolis effect. These modes are driven unstable by gravitational radiation reaction via the Chandrasekhar-Friedman-Schutz (CFS) mechanism \cite{C,FS}. In the absence of fluid dissipation, the CFS mechanism causes any mode that is retrograde in the co-rotating frame, but prograde in the inertial frame, to grow as gravitational radiation is emitted \cite{nils, Sharon}. The most unstable r-mode is the $n=3, m=2$ mode, called the r-mode throughout the rest of the paper (Here $n$ and $m$ label the degree and order of the Legendre functions associated with the mode.) The gravitational driving equals viscous dissipation for this mode along a critical curve in the angular velocity - temperature ($\Omega-T$) phase space.  Above the critical curve the r-mode is linearly unstable and the r-mode amplitude grows exponentially. Once the r-mode amplitude passes its first parametric instability threshold, other near-resonant inertial modes are excited via energy transfer from the r-mode. At this point nonlinear effects become important. The parametric instability threshold amplitude depends on internal neutron star physics and changes with angular velocity and temperature.  

 % If the star is close to the r-mode stability curve and the viscosity is an increasing function of temperature, oscillations around the stability curve can occur in which the star goes in and out of the unstable region. Otherwise, in the unstable regime, the r-mode grows exponentially until its amplitude becomes larger than the parametric instability threshold. At this point other near-resonant inertial modes in the star are excited via energy transfer from the r-mode and nonlinear effects become important. The parametric instability threshold depends on neutron star physics such as the strength of mode couplings and viscous dissipation and usually changes with angular velocity and temperature.
          
 %Young neutron stars are very dynamic systems. We take an initial temperature of $T = 10^{10}$ K and assume that the star is spinning close to break-up with angular velocity $\Omega \sim 0.67 \sqrt{\pi G \bar{\rho}}$, which places it well in the unstable regime. 
The first investigation that modeled the spin-down of a young neutron star due to the r-mode instability was performed by Lindblom {\it et al.\ }\cite{LOM} (See also Owen {\it et al.}~\cite{OwenEtAl} for a more detailed analysis.) They used a simple one-mode evolution model that assumes  the r-mode amplitude saturates because of nonlinear effects at some arbitrarily fixed value. The saturation amplitude was chosen to be of order 1. In this model, once the instability is saturated, the star spins down at fixed r-mode amplitude. Lindblom {\it et al.\ }estimated that a newborn neutron star would cool to approximately $10^9$ K and spin down from a frequency close to the Kepler frequency to about 100 Hz in $\sim$ 1 yr. In their calculation they included the effects of shear viscosity and bulk viscosity for ordinary neutron star matter composed of neutrons, protons and electrons and assumed modified URCA cooling.

Jones  \cite{Jones} and Lindblom \& Owen \cite{LO} pointed out that if the star contains exotic particles such as hyperons, internal processes could lead to a very high bulk viscosity in the cores of neutron stars. They predicted that for young neutron stars this viscosity would either eliminate the instability altogether or leave a short window of instability of up to a day or so for modified URCA cooling \cite{LO,mohit} that would render  the gravitational radiation undetectable. Andersson, Jones, and Kokkotas \cite{AJK} found that, in the case of strange stars, young neutron stars can evolve along the r-mode stability curve reaching a quasi-adiabatic equilibrium at low r-mode amplitudes. Similar conclusions were reached by Reisenegger \& Bonacic \cite{nonlinearbulk2} for large hyperon bulk viscosity. We include nonlinear effects and find that the spectrum of possible evolutions is more diverse. Quasi-equilibrium evolutions along the r-mode stability curve are just one scenario.

Schenk {\it et al.\ }\cite{Schenk} developed a formalism to study nonlinear interactions of the r-mode with other inertial modes. They assumed a small r-mode amplitude and treated the oscillations of the modes with weakly nonlinear perturbation theory via three-mode couplings. This assumption was tested by Arras {\it et al.\ } \cite{arras} and Brink {\it et al.} \cite{Jeandrew1, Jeandrew2, Jeandrew3,JeandrewThesis}. Arras 
{\it et al.} proposed that a turbulent cascade will develop in the strong driving regime. They estimated that the r-mode amplitude was small and could have values between $10^{-4} - 10^{-1}$. Brink {\it et al.\ }computed the interactions of about 5000 modes via approximately $1.3 \times 10^6$ couplings among modes with $n \le 30$. They modeled the star as incompressible and calculated the coupling coefficients analytically. The couplings were restricted to near resonant modes with a fractional detuning of $\delta \omega/(2 \Omega) < 0.002$.  Brink {\it et al.\ }showed that the nonlinear evolution saturates at amplitudes comparable with the lowest parametric instability threshold. They did not include spin or temperature evolution in their calculation.

In a previous paper \cite{us} we investigated the nonlinear saturation of the r-mode instability for neutron stars in LMXBs and included temperature and spin evolution. In this paper we begin a study of the nonlinear development of the r-mode instability in newborn neutron stars.  We use an effective three-mode treatment in that we include one triplet of modes with a statistically relevant coupling and detuning coefficient and treat it as the lowest parametric instability threshold. The triplet consists of the $n=3$, $m=2$ r-mode and two near-resonant inertial modes with $n=14$ and $n=15$ that couple to it.  The exact inertial modes that are excited when the r-mode grows above its parametric instability threshold will change as the star spins down. However, since the nearest resonance is statistically expected to be at a detuning of  $\delta \omega/2 \Omega \sim 10^{-4}$ and the mode coupling coefficient for the lowest parametric instability threshold is typically of order one~\cite{JeandrewThesis}, a mode triplet using these values should provide a qualitatively correct picture.

 The model also includes neutrino cooling via a combination of fast and slow processes, viscous heating due to hyperon bulk viscosity and boundary layer viscosity, and spin-down due to gravitational radiation and magnetic dipole radiation. In oder to explore possible nonlinear behaviors we vary: (1) the hyperon superfluid transition temperature $T_{\rm h}$, which is believed to be $\sim 10^9-10^{10}$ K; (2) the strength of the hyperon bulk viscosity;  (3) the boundary layer viscosity via the slippage factor $S_{\rm ns}$ that parametrizes the interaction between oscillating fluid core and the elastic crust \cite{LU, GA}; and (4) the fraction of the star that cools via direct URCA reactions $f_{\rm dU}$. 
 
 We find a variety of scenarios that depend on these parameters and on the initial r-mode amplitude. 
 The star cools until the cooling is balanced by viscous heating. It then follows a quasi-adiabatic evolution either on the r-mode stability curve or on other Cooling = Heating curves on which the neutrino cooling is balanced by viscous heating from the three modes. For low hyperon bulk viscosity or when we include only slow cooling, we find runaway behavior in which the energy dissipated by the two inertial modes is not sufficient to stop the growth of the r-mode amplitude and  several parametric instability thresholds are passed. Modeling such behavior accurately requires the inclusion of multiple mode triplets, which is beyond the scope of this paper.
       
The three mode problem is a natural first step in studying nonlinear effects for the r-mode instability. This model is valid as long as the r-mode amplitude does not grow
above parametric instability thresholds that are higher than the lowest threshold.
The thresholds are functions of the angular velocity and temperature of the star
and depend on the viscous damping rates of the modes and model details.
The star cools at constant angular velocity until the cooling is stopped by viscous
heating and then spins down oscillating between thermal equilibrium states until the
r-mode is no longer unstable.  The r-mode amplitudes we find are still fairly low
$\sim 10^{-2} - 10^{-3}$ and the spin-down torque due to gravitational radiation reaction
is typically lower than that due to magnetic dipole radiation. So, the spin down timescale, which
is approximately the timescale on which the r-mode is unstable, is dominated by the magnetic
dipole radiation timescale for $B \sim 10^{13}$ G.  Our evolutions are determined by
competitions between cooling and heating, gravitational driving and viscous damping, and
magnetic dipole and gravitational spin-down with the competition between cooling
and heating playing the most prominent role. We expect that some of this behavior
will hold for more sophisticated models as well.
 
 Our evolutions are determined by competitions between neutrino cooling and viscous heating, gravitational driving and viscous dissipation, neutrino cooling and gravitational driving, neutrino cooling and magnetic spin-down of the star with the competition between cooling and heating placing the most prominent role. The three-mode model is adequate as long as the mode amplitudes do not grow high enough to excite more modes in the star. Stable three-mode evolutions are more likely to happen for high viscosity such that due hyperons. %Depending on where in the $\Omega-T$ plane the neutrino cooling is stopped by viscous heating, the star spins down in different regions delimited by the r-mode stability curve.
 
The results are summarized in Sec.\ \ref{summary}. Model details are presented in Sec.\ \ref{MB}. Stable evolutions are examined in detail Sec.~\ref{stable}. The runaway evolutions are discussed in Sec.~\ref{unstable}. The prospects of detecting gravitational radiation are considered in Sec.~\ref{detection}. Limitations of the model are discussed in Sec.\ \ref{limitations}. Concluding remarks are presented in Sec.\ \ref{conclusion}. %Appendix A  compares timescales resulting from our method with those computed by Lockitch and Friedman \cite{KL} for bulk viscosity timescales.  Appendix B computes the mode dependence of boundary layer viscosity for the simple rigid crust proposed by Bildsten and Ushomirsky \cite{BU}. Appendix C derives the frequency change in the inertial modes due to the presence of a magnetic field. Appendix D contains a stability analysis of the evolution equations around thermal equilibrium and gives a rough explanation for thermal oscillations around the r-mode stability curve.
\section{Summary of Results}
\label{summary}
\begin{figure}
\begin{center}
%\leavevmode
%\epsfxsize=250pt
%\epsfbox{SchematicFigG.eps}
\epsfig{file = 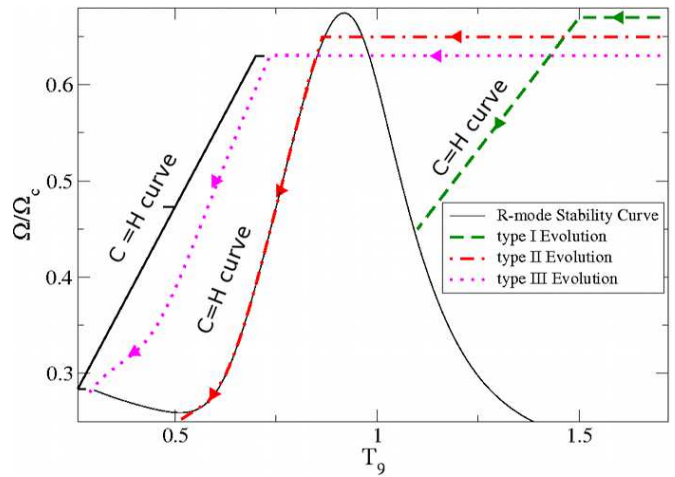, width = 250pt}
\caption{Schematic plot showing different trajectories that the star can follow in the $\tilde \Omega$ - $T_9$ plane. Here $\tilde \Omega = \Omega/\Omega_c$, $\Omega_c = \sqrt{\pi G \bar{\rho}}$, and $T_9 = T/(10^9 K)$. The Cooling = Heating ($C=H$) curves begin when the star starts spinning down and end when it reaches the stability curve. For simplicity we did not include mixed type I-II and type II-III evolutions, where the thermal equilibrium becomes unstable and the star cools from one $C=H$ curve to another. A flow chart that includes all possible scenarios is displayed in Fig.\ \ref{diagram}. }
%placeholder
\label{schematic}
\end{center}
\end{figure}
 \begin{figure*}[ht]
  \begin{center}
\epsfig{file = 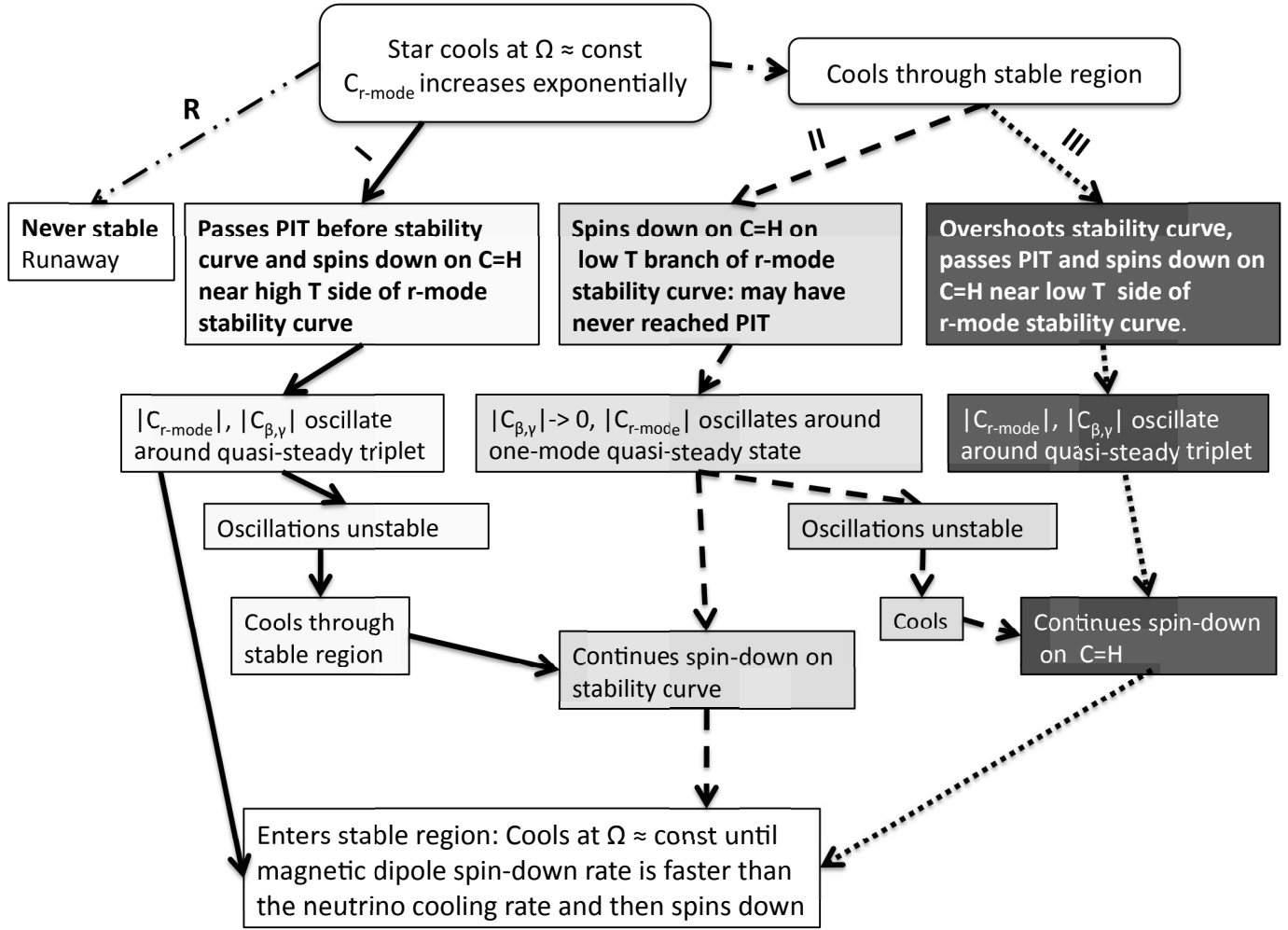, angle = - 90, width = 20 cm}
\end{center}
\caption{Flow diagram for the various evolution scenarios. The parametric instability threshold is abbreviated as PIT.}
 \label{diagram}
\end{figure*}  
%The Cooling = Heating ($C=H$) curves begin when the star starts spinning down and end when it reaches the stability curve. This would be true only for a quasi-steady evolution. For an evolution that does not assume quasi-stationary amplitudes, the star oscillates around the quasi-steady solution
 Our evolutions start with an initial temperature of  $10^{10}$ K and an initial rotation rate at the Kepler break-up frequency  $\Omega = 0.67 \sqrt{\pi G \bar{\rho}}$, where $\bar{\rho}$ is the average density of the star.  These initial conditions place the neutron star above the r-mode stability curve in the unstable region. The evolutions are independent of the initial temperature for temperatures for which the cooling rate is faster than the gravitational driving rate.  The star cools at approximately constant angular velocity until the cooling is stopped by viscous heating. Once thermal equilibrium is reached, the star spins down oscillating though a series of thermal equilibrium states.  The spin-down is the longest part of the evolution lasting at least several years, while the initial cooling at constant angular velocity lasts minutes to hours depending on whether low or fast cooling processes dominate. So, the timescale on which the r-mode instability is active and gravitational radiation is emitted is approximately the spin-down timescale.
 
   The spin-down part of the trajectory of the star can be approximated by a Cooling = Heating ($C=H$) curve. This curve is not unique and is determined by equating the neutrino cooling and the viscous heating using quasi-stationary amplitude solutions (Eqs.\ (\ref{stationarySol}) or Eq.\ (\ref{OneModeQuasi})). Our schematic hyperon bulk viscosity model corresponds to an r-mode stability curve with one maximum (See Fig.\ \ref{schematic}).  We find three possible $C=H$ curves: one on each side of the maximum, and one that is on the r-mode stability curve, where gravitational driving equals viscous dissipation. The latter corresponds to evolutions for which a one-mode model is adequate and the former two $C=H$ curves approximate triplet evolutions. 
 
 The trajectory that the star follows depends on whether the r-mode amplitude passes its parametric instability threshold before the r-mode stability curve or afterwards. If the parametric instability threshold is reached first, the star will spin down on the $C=H$ near the high T side ($T> T_{\rm peak}$) of the r-mode stability curve ({\bf type I} evolution). In this scenario, the viscous heating due to all three modes  balances the cooling. If the r-mode stability curve is reached first, the star can spin down following the r-mode stability curve ({\bf type II} evolution) or, for lower initial r-mode amplitude, cool through that thermal equilibrium region and spin down on the $C=H$ near the low $T$ side ($T < T_{\rm peak}$) of the r-mode stability curve ({\bf type III} evolution). As the star spins down and cools slowly, the thermal equilibrium can become unstable. When this happens the star cools to the next closest thermal equilibrium region leading to mixed {\bf type I-II} (See Fig.\ \ref{A1} for an example.) and {\bf type II-III} (See Fig.\ \ref{Tch12}.) evolutions.
 
The quasi-stationary solutions are independent of initial conditions. Evolutions in which the three modes are sufficient to model the behavior follow trajectories that oscillate around $C=H$ curves. So, they are roughly independent of initial conditions as well. However, the initial r-mode amplitude can determine which type of evolution will occur. At a given viscosity, cooling and initial spin-frequency, for initial r-mode amplitudes above a certain $C_\alpha(0)$ the r-mode amplitude will always reach parametric instability before the stability curve and hence a type I evolution will occur. And below this value type II evolutions will occur. A similar amplitude threshold exists between type II and type III evolutions.
  
  If no thermal equilibrium can be reached, the r-mode amplitude grows above several parametric instability thresholds and excites more inertial modes. In order to accurately model runaway behavior more mode triplets need to be included. This is beyond the scope of this paper. We obtain runaway scenarios by either lowering the hyperon bulk viscosity or by turning off the fast neutrino cooling. So, for low mass neutron stars that have lower central densities and no hyperon population, multiple triplets of modes are required to model the nonlinear behavior accurately. If energy transfer to inertial modes does not stop the instability at a low r-mode amplitude,  other nonlinear saturation effects would become important such as nonlinear bulk viscosity \cite{nonlinearbulk2}. 
 
  The evolutionary scenarios described here are summarized in Fig.\ \ref{diagram}. Each stable evolution scenario is presented in more detail in Sec.\ \ref{stable} and the unstable evolutions are discussed in Sec.\ \ref{unstable}. We consider an evolution to be 'stable' when one mode triplet is adequate to model the evolution. In this case the neutrino cooling is stopped by viscous heating due to the three modes and the star spins down oscillating between thermal equilibrium states. 'Unstable' or 'runaway' evolutions occur when the star reaches thermal equilibrium only after the r-mode and the inertial modes overshoot several parametric instability thresholds reaching amplitudes of order unity. Modeling such scenarios accurately requires a larger mode network.
  
 \section{Setup}
\label{MB}
\subsection{Three-mode Evolution Equations}
The evolution equations for the mode amplitudes when the $n=3, m=2$ r-mode couples to two near-resonant inertial modes can be written as 
\begin{eqnarray}
\label{Amplitudes}
\frac{d C_\alpha}{d\tau} &=& i \tilde{\omega}_\alpha C_\alpha + \frac{\gamma_\alpha}{\Omega} C_\alpha - \frac{2 i \tilde{\omega}_\alpha \tilde{\kappa}}{\sqrt{\Omega}} C_\beta C_\gamma , \\ \nonumber
\frac{d C_\beta}{d\tau} &=& i \tilde{\omega}_\beta C_\beta - \frac{\gamma_\beta}{\Omega} C_\beta - \frac{2 i \tilde{\omega}_\beta \tilde{\kappa}}{\sqrt{\Omega}} C_\alpha C_\gamma^\star , \\ \nonumber
\frac{d C_\gamma}{d\tau} &=& i \tilde{\omega}_\gamma C_\gamma - \frac{\gamma_\gamma}{\Omega} C_\gamma - \frac{2 i \tilde{\omega}_\gamma \tilde{\kappa}}{\sqrt{\Omega}} C_\alpha C_\beta^\star .
\end{eqnarray}
The scaled frequency $\tilde{\omega}_j$ is $\tilde{\omega}_j = \omega_j/\Omega$, the dissipation rates of the inertial modes, also called daughter modes, are $\gamma_\beta$ and $\gamma_\gamma$, $\gamma_\alpha$ is the sum of the driving and damping rates of the r-mode  $\gamma_\alpha = \gamma_{GR} - \gamma_{\alpha\, v}$, and the dimensionless coupling is $\tilde{\kappa} = \kappa/(M R^2 \Omega^2)$.   The rotational phase $\tau$ is defined by $d \tau =\Omega\;dt$. In terms of the amplitude variables of Schenk {\it et al.}  \cite{Schenk}  and Brink {\it et al.} \cite{Jeandrew1,Jeandrew2,Jeandrew3} $C_j = \sqrt{\Omega(t)} c_j(t)$, which are normalized to unit energy, i.e., mode amplitudes of $c_j = 1$ correspond to a mode energy $E_j = \epsilon = M R^2 \Omega^2$  equal to the rotational energy of the star$^2$.  The equations of motion in Schenk {\it et al.} assume constant $\Omega$. We are interested in a situation in which the uniform angular velocity of the star changes very slowly on a timescale of a rotation period. Writing the equation of motion in terms of rotational phase and the amplitude variables $C_j$ removes the time dependence (Without rescaling, the amplitude evolution equations would have terms proportional to $d{\Omega}/dt$.) A derivation of the equations of motion for the three mode system from the Lagrangian density can be found in Appendix A of Ref. \cite{us}.

\footnotetext[2]{The conversion between the amplitudes of Schenk et al. and the $\alpha$ used by Lindblom et al. \cite{LOM} is $c_\alpha  = \sqrt{\tilde J/2}\; \alpha \approx 0.1 \alpha$ for an $n=1$ polytrope  with $\tilde J = 1/(M R^4) \int_0^R dr \rho r^6$.}
%A derivation of the equations of motion for the three-mode system in the limit of slow rotation in terms of rotational phase $\tau$  ($d \tau =\Omega\;dt$) from the Lagrangian density can be found in Appendix A of Ref. \cite{us}.

We further rescale the equations:
\begin{eqnarray}
\label{eqcode}
\frac{d \bar{C}_\alpha}{d\tilde{\tau}} &=& \frac{i \tilde{\omega}_\alpha}{|\delta \tilde{\omega}|} \bar{C}_\alpha + \frac{\tilde{\gamma}_\alpha}{|\delta \tilde{\omega}| \tilde{\Omega}} \bar{C}_\alpha - \frac{i}{2 \sqrt{\tilde{\Omega}}} \bar{C}_\beta \bar{C}_\gamma ,  \\ \nonumber
\frac{d \bar{C}_\beta}{d \tilde{\tau}} &=& \frac{i \tilde{\omega}_\beta}{|\delta \tilde{\omega}|} \bar{C}_\beta - \frac{\tilde{\gamma}_\beta}{|\delta \tilde{\omega}| \tilde{\Omega}} \bar{C}_\beta - \frac{i}{2 \sqrt{\tilde{\Omega}}} \bar{C}_\alpha \bar{C}_\gamma^\star , \\ \nonumber
\frac{d \bar{C}_\gamma}{d\tilde{\tau}}&=&\frac{i \tilde{\omega}_\gamma}{|\delta \tilde{\omega}|} \bar{C}_\gamma - \frac{\tilde{\gamma}_\gamma}{|\delta \tilde{\omega}| \tilde{\Omega}} \bar{C}_\gamma - \frac{i}{2 \sqrt{\tilde{\Omega}}} \bar{C}_\alpha \bar{C}_\beta^\star,
\end{eqnarray}  

\noindent where we have used the same rescaling as in Ref.\ \cite{us} with $\bar{C}_j  = C_j/|C_j|_0$, $\tilde{\gamma}_j = \gamma_j/\Omega_c$, and $\Omega_c$ is a fixed  angular frequency chosen for reference.

The rotational phase $\tau$ is rescaled by the fractional detuning as $\tilde{\tau} = \tau |\delta \tilde{\omega}|$ and the mode amplitudes scaled by the zero-viscosity parametric instability threshold
\begin{eqnarray}
\label{norm}
|C_\alpha|_0 &=& \frac{|\delta \tilde{\omega}| \sqrt{\Omega_c}}{4 \tilde{\kappa} \sqrt{\tilde{\omega}_\beta \tilde{\omega}_\gamma}}, \; |C_\beta|_0 = \frac{|\delta \tilde{\omega}| \sqrt{\Omega_c}}{4 \tilde{\kappa} \sqrt{\tilde{\omega}_\alpha \tilde{\omega}_\gamma}}, \\ \nonumber
 |C_\gamma|_0 &=& \frac{|\delta \tilde{\omega}| \sqrt{\Omega_c}}{4 \tilde{\kappa} \sqrt{\tilde{\omega}_\beta \tilde{\omega}_\alpha}} \;.
\end{eqnarray}
% {\bf This rescaling allows us to work with amplitude variables that are of order unity during most of the evolution.}
\subsection{Driving and Damping Rates}
\label{DD}
For our benchmark calculations, we adopt the neutron star model of Owen {\it et al.} Ref. \cite{OwenEtAl}  ($n=1$ polytrope, $M = 1.4 M_\odot$, $\Omega_c = 8.4 \times 10^3 \; \rm{rad} \; \rm{sec}^{-1}$ and $R = 12.5$ km) and use their gravitational driving rate for the r-mode
\begin{eqnarray}
\gamma_{GR}(\Omega) &\simeq& 0.05\,{\rm sec^{-1}} M_{1.4} R_{12.5}^4\left(\frac{\nu}{\rm{1 \, kHz}}\right)^6 \\ \nonumber
&\simeq& \frac{\tilde{\Omega}^6}{3.26} \; \rm{sec}^{-1},
\end{eqnarray}
where $M_{1.4} = M/1.4 M_\odot$, $R_{12.5}  = R/12.5\, \rm{km}$ and $\tilde \Omega = \Omega/\Omega_c$.
If hyperons are present in the core of the neutron star then hyperon bulk viscosity dominates other forms of bulk viscosity. In this paper we consider hyperon bulk viscosity and boundary layer viscosity to be the main sources of  dissipation. We use a schematic model with a few adjustable parameters. See Appendix B for a detailed description.
 
\subsection{Angular Momentum and Temperature Evolution}
Angular momentum is lost via gravitational wave emission
\begin{equation}
\label{consAng}
\frac{dJ}{dt} = 2  \gamma_{GR} J_{c\;\rm{rmode}} - \frac{I \Omega}{\tau_M}, 
\end{equation}
where $J_{c\;\rm{rmode}} = - (m_\alpha/\omega_{\alpha}) \epsilon_\alpha |c_\alpha|^2 = - 3 M R^2 \Omega |c_\alpha|^2 = - 3 M R^2 |C_\alpha|^2$, $I$ is the moment of inertia of the star, $I \Omega/\tau_M$ is the magnetic breaking torque, and $\tau_M$ is the corresponding timescale. We have adopted the simplest magnetic dipole model with a timescale
\begin{eqnarray}
\frac{1}{\tau_M} &=& \frac{\mu^2 \Omega^2}{6 c^3 I}  = \frac{1}{3 \times 10^9 \; {\rm sec}} \frac{\mu_{30}^2 \nu_{\rm kHz}^2}{M_{1.4} R_{10}^2} \\ \nonumber
&=& \frac{B_{12}^2 \tilde{\Omega}^2 R_{12.5}}{6.7 \times 10^8}  \; {\rm sec}^{-1},
%\frac{1}{1.2 \times 10^9 \rm sec^{-1}} \frac{R_{12.5}^4 B_{12}^2}{M_{1.4}} \left(\frac{\nu}{\rm 1 kHz}\right)^2 \qquad \\ \nonumber
%&=&  \frac{B_{12}^2 \tilde \Omega^2}{6.7 \times 10^8 \rm sec^{-1}}, 
\end{eqnarray}
where $B_{12} = B/(10^{12}$ G) and the magnetic dipole moment $\mu =  10^{30} \mu_{30}$ G cm$^3$. Fallback disks observed around isolated highly magnetic neutron stars \cite{Axp} and gas inside the magnetosphere could make the spin-down more complicated.  

Eq.\ (\ref{consAng}) can be rewritten in terms of the scaled variables in Eq.\ (\ref{norm}) as
\begin{equation}
\label{ang2}
\frac{dJ}{d\tilde{\tau}} = -  \frac{6 \tilde{\gamma}_{\rm{GR}}}{\tilde{\Omega}} \frac{M R^2 \Omega_c |\delta \tilde{\omega}|}{(4 \tilde{\kappa})^2 \tilde{\omega}_\beta \tilde{\omega}_\gamma}  |\bar{C}_\alpha|^2  - \frac{I}{|\delta \tilde \omega| \tau_M}.
\end{equation}
We then consider the thermal evolution of the star
\begin{eqnarray}
\label{consE}
C(T) \frac{dT}{dt} &=& \sum_j 2 E_j  \gamma_j  - L_\nu(T) \\ \nonumber
&=& 2 M R^2 \Omega (\gamma_{\alpha\,v} |C_\alpha|^2 + \gamma_\beta |C_\beta|^2 \\ \nonumber
 &+& \gamma_\gamma |C_\gamma|^2)  - L_\nu(T).
\end{eqnarray} 
We use a specific heat $C(T) = 1.5 \times 10^{39} T_9$~erg K$^{-1}$ \cite{Max}, which assumes normal neutrons. Yakovlev and Pethick \cite{YP} found that cooling observations are consistent with weak neutron superfluidity with critical temperatures of $T_n \le 2\times 10^8$ K and strong proton superfluidity. The energy gap calculations for neutron superfluidity are still very uncertain. Our temperatures are typically above $5\times 10^8$ K and we choose to ignore neutron superfluidity.  If neutrons are superfluid, the heat capacity is reduced. A lower bound to $C(T)$ is the electron contribution, which is about 20 times lower than the neutron-neutron specific heat used here \cite{YP2}.

 The neutrino luminosity is a superposition of fast and slow processes
\begin{eqnarray}
L_\nu &=& L_{\rm{dU}} T_9^6 R_{\rm{dU}}(T/T_p) + L_{\rm{mU}} T_9^8 R_{\rm{mU}}(T/T_p) \\ \nonumber
 &+& L_{\rm{e-i}} T_9^6 + L_{\rm{n-n}} T_9^8,
\end{eqnarray}
where the proton superfluid reduction factors for the modified and direct URCA reactions are taken from  Eqs. (32) and (51) of Ref. \cite{Yakovlev2}. The other constants are adopted from Ref.\ \cite{Wagoner}:   $L_{\rm{mU}} = 1.0 \times 10^{40} \; \rm{erg\; sec}^{-1}$, $L_{\rm{dU}} = f_{\rm{dU}} 
\times 10^{46} \; \rm{erg\; sec}^{-1}$, and the electron-ion and neutron-neutron neutrino bremsstrahlung are given by  $L_{\rm{e-i}} = 9.1 \times 10^{35}\;\rm{erg\; sec}^{-1}$,  $L_{\rm{n-n}} \approx 10^{38} \; \rm{erg\; sec}^{-1}$ \cite{Yakovlev, Yakovlev2,YP2}. The fraction of the star $f_{\rm{dU}}$ that is above the density threshold for direct URCA reactions is in general dependent on the equation of state \cite{YP}. We take it to be $10 \%$ for all stable evolutions in this paper.
%\begin{eqnarray}
%R_{\rm{dU}}(T/T_p) &=&\left[0.2312 + \sqrt{(0.76880)^2 + (0.1438 v)^2}\right]^{5.5} \;\;\; \\ \nonumber
% &\times& \exp\left(3.427- \sqrt{(3.427)^2 + v^2}\right) , \\ \nonumber
%R_{\rm{mU}}(T/T_p) &=& \left(0.2414 + \sqrt{(0.7586)^2+(0.1318 v)^2}\right)^7 , \\ \nonumber
%&\times&\exp\left(5.339-\sqrt{(5.339)^2+(2 v)^2}\right)
%\end{eqnarray}
%where  the dimensionless gap amplitude $v$ for the singlet type superfluidity is given by
%\begin{eqnarray}
%a=0.1477+\sqrt{(0.8523)^2+(0.1175 v)^2} \\ \nonumber
%b=0.1477+\sqrt{(0.8523)^2+(0.1297 v)^2}
%\end{eqnarray}
%\begin{equation}
%v=\sqrt{1-\frac{T}{T_p}} \left(1.456-0.157 \sqrt{\frac{T_p}{T}} + 1.764 \frac{T_p}{T} \right).
%\end{equation}
We use a constant $T_p = 5.0 \times 10^9$ K, which is consistent with Fig.~5 of \cite{YP}. A more realistic model would involve a critical temperature that changes with the density of the star.

In terms of the scaled variables Eq.\ (\ref{consE}) becomes
\begin{eqnarray}
\label{thermalevol}
C(T) \frac{dT}{d\tilde{\tau}} = \frac{2 M R^2 \Omega_c^2 |\delta \tilde{\omega}|}{(4 \tilde{\kappa})^2 \tilde{\omega}_\alpha  \tilde{\omega}_\beta  \tilde{\omega}_\gamma} (\tilde{\omega}_\alpha \tilde{\gamma}_{\alpha\,v} |\bar{C}_\alpha|^2 + \tilde{\omega}_\beta \tilde{\gamma}_\beta |\bar{C}_\beta|^2  \\ \nonumber
{}+ \tilde{\omega}_\gamma \tilde{\gamma}_\gamma |\bar{C}_\gamma|^2)
-  \frac{L_\nu(T)}{ \Omega_c \tilde{\Omega} |\delta \tilde{\omega}|}.
\end{eqnarray}

\subsection{Quasi-Adiabatic Spin and Temperature Evolution}
In type I and type III evolutions, after an initial precursor, the mode amplitudes settle in and later oscillate around their quasi-stationary solutions
\begin{eqnarray}
\label{stationarySol}
|\bar{C}_\alpha|^2 = \frac{4 \tilde{\gamma}_\beta \tilde{\gamma}_\gamma}{\tilde{\Omega} |\delta \tilde{\omega}|^2} \left(1 + \frac{1}{\tan^2 \phi}\right) , \\ \nonumber
|\bar{C}_\beta|^2 = \frac{4 \tilde{\gamma}_\alpha \tilde{\gamma}_\gamma}{\tilde{\Omega} |\delta \tilde{\omega}|^2} \left(1 + \frac{1}{\tan^2 \phi}\right) , \\ \nonumber
|\bar{C}_\gamma|^2 = \frac{4 \tilde{\gamma}_\alpha \tilde{\gamma}_\beta}{\tilde{\Omega} |\delta \tilde{\omega}|^2} \left(1 + \frac{1}{\tan^2 \phi}\right) , \\ \nonumber
\tan \phi = \frac{\tilde{\gamma}_\beta + \tilde{\gamma}_\gamma - \tilde{\gamma}_\alpha}{\tilde{\Omega} |\delta \tilde{\omega}|} .
\end{eqnarray}
Appendix A provides a brief derivation. 

 In the limit of $|\delta \omega|/\gamma_j << 1$, the quasi-steady mode amplitudes simplify to
\begin{eqnarray}
\label{TripletQuasi}
|\bar{C}_\alpha|^2 \approx \frac{4 \tilde{\gamma}_\beta \tilde{\gamma}_\gamma}{\tilde{\Omega} |\delta \tilde{\omega}|^2}, \;
|\bar{C}_\beta|^2 \approx \frac{4 \tilde{\gamma}_\alpha \tilde{\gamma}_\gamma}{\tilde{\Omega} |\delta \tilde{\omega}|^2}, \;
|\bar{C}_\gamma|^2 \approx \frac{4 \tilde{\gamma}_\alpha \tilde{\gamma}_\beta}{\tilde{\Omega} |\delta \tilde{\omega}|^2}. \\ \nonumber
\end{eqnarray}
In the limit when  $|\delta \omega|/\gamma_j >> 1$
\begin{eqnarray}
\label{TripletQuasi2}
|\bar{C}_\alpha|^2 &\approx& \frac{4 \tilde{\gamma}_\beta \tilde{\gamma}_\gamma \tilde{\Omega}}{(\tilde\gamma_\beta + \tilde \gamma_\gamma - \tilde \gamma_\alpha)^2}, \;
|\bar{C}_\beta|^2 \approx \frac{4 \tilde{\gamma}_\alpha \tilde{\gamma}_\gamma \tilde{\Omega}}{(\tilde\gamma_\beta + \tilde \gamma_\gamma - \tilde \gamma_\alpha)^2},\;\;\;\;\;\;\;\;\; \\ \nonumber
|\bar{C}_\gamma|^2 &\approx& \frac{4 \tilde{\gamma}_\alpha \tilde{\gamma}_\beta \tilde{\Omega}}{(\tilde\gamma_\beta + \tilde \gamma_\gamma - \tilde \gamma_\alpha)^2}. \\ \nonumber
\end{eqnarray}

Assuming that $dC_i/d\tilde{\tau} \approx 0$, $J\approx I \Omega$ and using Eqs.~(\ref{stationarySol}) in Eqs.\ (\ref{ang2}-\ref{thermalevol}) the spin and thermal evolution equations can be rewritten as
\begin{equation}
\label{spinSteady}
\frac{d \tilde{\Omega}}{d\tilde{\tau}} = - \frac{6 \tilde{\gamma}_{\rm{GR}}}{\tilde{\Omega}^2 |\delta \tilde{w}|} \frac{\tilde{\gamma}_\beta \tilde{\gamma}_\gamma}{4 \tilde{\kappa}^2\tilde{I}  \tilde{\omega}_\beta \tilde{\omega}_\gamma} \left(1 + \frac{1}{\tan^2 \phi}\right) - \frac{1}{\tau_M \Omega_c |\delta \tilde \omega|},
\end{equation}
where $\tilde{I} = I/(M R^2)$, and
\begin{eqnarray}
\label{thermalSteady}
C(T) \frac{dT}{d\tilde{\tau}} &=&\frac{2 M R^2 \Omega_c^2}{4 \tilde{\kappa}^2 \tilde{\omega}_\alpha  \tilde{\omega}_\beta  \tilde{\omega}_\gamma} \frac{\tilde{\gamma}_\alpha \tilde{\gamma}_\beta \tilde{\gamma}_\gamma}{\tilde{\Omega}  |\delta \tilde{\omega}|}  \left(1 + \frac{1}{\tan^2 \phi}\right) \qquad \\ \nonumber 
&& \times \left(\frac{\tilde{\omega}_\alpha \tilde{\gamma}_{\alpha,v}}{\tilde{\gamma}_\alpha}+ \tilde{\omega}_\beta +
\tilde{\omega}_\gamma \right) - \frac{L_\nu(T)}{\Omega_c \tilde{\Omega} |\delta \tilde{\omega}|}.
\end{eqnarray}
By setting the right hand side of the above equation to zero, one can find the Cooling = Heating ($C=H$) curve. Below, we find that Eqs.\ (\ref{spinSteady})-(\ref{thermalSteady}) approximate well the spin-down part of the evolution for type I and III scenarios throughout the stable regime. 

In type II evolutions the trajectory settles on or oscillates around the low T branch of the r-mode stability curve. The thermal equilibrium on the stability curve corresponds to zero daughter mode amplitudes and an r-mode amplitude determined by the one-mode quasi-stationary state 
\begin{equation}
|\bar{C}_\alpha|^2 =|\bar{C}_{\rm C=H \, one-mode}|^2 = \left(\frac{L_\nu}{\tilde \gamma_{\alpha\,v}}\right) \frac{(4 \tilde \kappa)^2 \tilde \omega_\beta \tilde \omega_\gamma}{2 M R^2 \Omega_c^3 \tilde \Omega |\delta \tilde \omega|^2}.
\label{OneModeQuasi}
\end{equation}
%In the strong damping limit the angular velocity at which the magnetic dipole spin-down is equal to the gravitational spin-down torque
%\begin{equation}
%\Omega_{MD = GR} = 2 \tilde \kappa B_{13} \left(\frac{3.26\; \rm{sec}}{6.7 \times 10^6\;\rm{sec}}\right)^{1/2} \left(\frac{ \tilde I \tilde \omega_\beta \tilde \omega_\gamma}{6 \tilde \gamma_\beta \tilde \gamma_\gamma}\right)^{1/2}.
%\end{equation}
%In the $|\delta \omega|/\gamma_j >> 1$ limit
%\begin{eqnarray}
%&&\Omega_{MD = GR} \left(\frac{\tilde{\gamma}_\beta \tilde{\gamma}_\gamma}{(\tilde\gamma_\beta + \tilde \gamma_\gamma - \tilde \gamma_\alpha)^2}\right)^{1/4} = \left(\frac{3.26}{6.7 \times 10^6}\right)^{1/4} \\ \nonumber
%&& \times \left(\frac{B_{13}}{|\delta \tilde \omega|}\right)^{1/2} \left(\frac{4 \tilde \kappa^2 \tilde I \tilde \omega_\beta \tilde \omega_\gamma}{6}\right)^{1/4}
%\end{eqnarray}

\subsection{Validity of Approximations}
\label{validity}
\subsubsection{Density of Resonances and the Effective Three Mode Coupling Approximation}
\begin{figure}[h]
\begin{center}
\leavevmode
\epsfxsize=250pt
\newline \newline\newline
\epsfbox{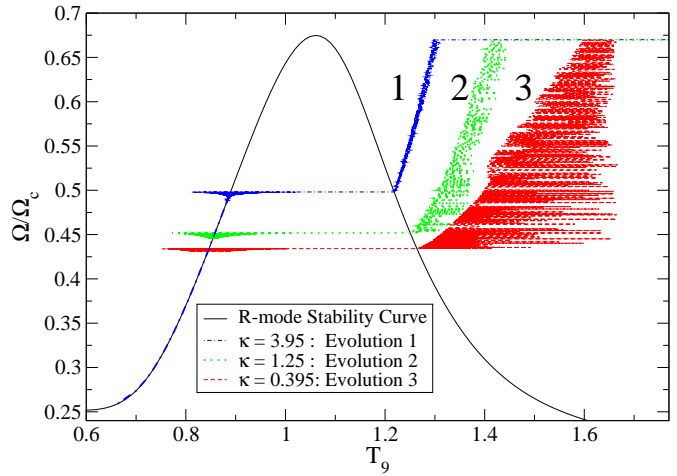}
\caption{The trajectories in the $\tilde\Omega - T_9$ plane for evolutions with different coupling coefficients are shown for three values of $\kappa$: $3.95$, $1.25$ and $0.395$. All other parameters are the same. The hyperon superfluidity temperature is $T_{\rm h} = 6.0 \times 10^9$ K. The cooling is stopped by the heating later for larger $\kappa$ and the thermal oscillations are smaller.}
%placeholder
 \renewcommand{\arraystretch}{0.75}
 \renewcommand{\topfraction}{0.6}
\label{DiffKappa}
\end{center}
\end{figure}

We choose a mode triplet with detuning, coupling coefficient, and viscous damping rates that were found to be 
representative for the first near-resonance of a dense set of modes confined to the 
$[- 2 \Omega, 2 \Omega]$ frequency range. Brink (See Chapter 5.2 of Ref.\ \cite{JeandrewThesis}) estimated statistically the expected lowest detuning for modes that are uniformly distributed. She considered all modes that couple to the r-mode and found  a detuning of $\delta \omega/(2 \Omega) \sim 10^{-4}$ for $n\le 14$. For a uniform frequency distribution this lowest detuning is inversely proportional to the number of direct couplings to the r-mode $N$, which increases drastically with $n$ ($N \approx n_{\rm max}^4/6$ for $n \le n_{\rm max}$). However, the viscosity also increases with $n$. Hyperon bulk viscosity dominates at $T \sim 10^9$ K, and we estimate that it scales roughly as $\gamma_{\rm hb} \propto n^2$. A very rough upper bound of $n \sim 100$ can be obtained for  $\gamma_{\rm hb} \sim \omega$. The exact $n$ cutoff beyond which higher order modes can be ignored is model dependent, and finding it involves an extensive calculation of damping rates, which is left for future investigation.

 Preliminary calculations \cite{SharonPrivate} show that when second order corrections to the mode frequencies are included, the average lowest detuning is still of order $\delta \omega/(2 \Omega) \sim 10^{-4}$.  The  exact $n$ and $m$ values of the modes that have this low detuning change with angular velocity. So, in principle, a more accurate modeling of the spin-down process would include multiple mode triplets and allow the identities of the modes comprising the lowest parametric instability threshold to change as the star evolves. However, since the lowest resonance always occurs for similar high $n$ modes with damping rates varying within a factor of 2, we conjecture that one mode triplet with these typical coupling  $\kappa \sim 1$ and detuning coefficients can act as an `effective' lowest resonance.

 To test the dependence of our evolutions on the coupling coefficient we vary it  by factors of $\sqrt{10}$ between $\kappa = 0.395$ and $\kappa = 3.95$, while keeping all other parameters the same. Fig.\ \ref{DiffKappa} shows that the evolutions change but are qualitatively similar.  Increasing $\kappa$ decreases the parametric instability threshold and the quasi-stationary mode amplitudes. This lowers the viscous heating due to the three modes. The cooling is stopped later increasing the angular velocity at which the star intersects the r-mode stability curve. % increases with $\kappa$. 
The system is less dynamic for larger $\kappa$ and the amplitudes of the thermal oscillations are lower. We also varied the detuning by a factor of 100, between $10^{-6}$ and $10^{-4}$, and did not observe significant changes in the evolution.  The exact temperature at which the star spins down and the initial r-mode amplitude $|C_\alpha|(0)$ at which the transition between different types of evolutions occurs change with $\kappa$ and $\delta \omega$, but, overall,  the evolution scenarios we obtain are qualitatively robust.
%This occurs because for type I evolutions the damping is strong enough $\delta \omega/(\gamma_\beta + \gamma_\gamma) << 1$ to make the parametric instability threshold insensitive to the value of the detuning $|C_\alpha|_{\rm PIT} \propto \sqrt{\gamma_\beta \gamma_\gamma}/\kappa$. In type II evolutions the daughter modes do not play a significant role and the r-mode amplitude settles or oscillates around the one-mode quasi-stationary solution. So, changing the detuning or coupling coefficient does not affect the evolution significantly. In type III evolutions changing the detuning and $\kappa$ moves the $C=H$ curve on which the star spins down. The initial r-mode amplitude $|C_\alpha|(0)$ for which the transition between type I and type II evolutions and between type II and type III evolutions occurs also changes.
 
 A shortcoming of the three-mode model is that there are cases when the r-mode grows above several parametric instability thresholds and the system runs away.  In order to model this scenario accurately one would have to include an oscillator network that takes into account not only direct couplings to the r-mode, but also couplings among inertial modes similar to the work of  Brink {\it et al.} \cite{Jeandrew1, Jeandrew2, Jeandrew3}.  %Such simulations are beyond the scope of this paper. 
 
\subsubsection{Buoyancy Effects}
 \label{buoyancy}
 
 Buoyancy effects are expected to be important when the angular velocity of the star is comparable with
 the Brunt-$\rm{V\ddot{a}is\ddot{a}l\ddot{a}}$ frequency. In a recent paper, Passamonti {\it et al.} \cite{AP} compared the inertial modes of a non-stratified star with the g-modes of a stratified system.  They found that each g-mode they studied approached a particular inertial mode as the rotation rate was increased. The $n=3$, $m=2$ r-mode does not have a g-mode counterpart. The conjecture is that if composition gradients are included, inertial modes will approach g-modes at low rotation rates, when buoyancy dominates. As $\Omega$ decreases the inertial mode frequencies level off rather than decreasing $\propto \Omega$. Once this happens, the r-mode loses its resonant modes, and at low rotation rates the r-mode amplitude would grow unabated as long as gravitational driving dominates viscous dissipation.
 
The Brunt-$\rm{V\ddot{a}is\ddot{a}l\ddot{a}}$ angular frequency is
  \begin{equation}
N \equiv \sqrt{- \frac{g}{\rho} \left(\frac{\partial \rho}{\partial x} \right)_P \frac{dx}{dr}} \sim   \sqrt{G \rho x} \sim 10^3 \; \rm{rad/sec},
 \end{equation}
 where the proton fraction is taken to be $x \simeq 5.6 \times 10^{-3} (\rho/\rho_{\rm nuc})$ and
the nuclear density $\rho_{\rm nuc} = 2.8 \times 10^{14}\; \rm{g cm^{-3}}$ \cite{Lai1999}. This corresponds to a spin frequency of 
\begin{equation}
\nu_N \approx 160 \;\rm{Hz}
\end{equation}

When $2 \Omega$ is of the order $N$ or lower the spectrum of inertial modes will change and the r-mode will lose it near-resonant modes. However, at $\Omega \approx N/2$ or $\nu \approx 80 $ Hz, the r-mode is in the stable regime for $S_{\rm ns} \ge 0.01$ for our boundary layer viscosity model, and hence this window of no nonlinear effects could be irrelevant.   %$N = 2\Omega$ would correspond to an angular frequency of the star of $\Omega = 500\; \rm{rad \, sec^{-1}}$ and a spin frequency $\nu \approx 80$ Hz.

For our viscosity model for a critical temperature of $T_h = 1.2 \times 10^{10}$ K, the minimum spin frequency on the r-mode stability curve for $T<T_{\rm peak}$ and $S_{\rm ns} \approx 0.3$ is at $\nu_{\rm min} \approx 300$ Hz ($\Omega \approx 0.22 \sqrt{\pi G \bar{\rho}}$).  Passamonti {\it et al.} find that Coriolis effects dominate beyond $\tilde{\Omega} \sim 0.3$ \cite{AP}. So, ignoring buoyancy in our evolutions should not lead to large qualitative changes.  This frequency scales as
\begin{equation}
\nu_{\rm min} \approx 300 \,{\rm Hz}\, \left(\frac{S_{\rm ns}}{0.3}\right)^{4/11} \left(\frac{T_h}{1.2 \times 10^{10} \, \rm K}\right)^{-2/11},
\label{ommin}
\end{equation}
when the temperature at the minimum is much smaller than the critical temperature for hyperons $T_{\rm min} << T_h$. It is important to note that $\nu_{\rm min}$ is not very sensitive to the critical temperature $T_h$, but moderately sensitive to the slippage factor $S_{\rm ns}$. At $T_h = 1.2 \times 10^{10}$ K we find: $\nu_{\rm min} \approx 90$ Hz  for $S_{\rm ns} = 0.01$, $\nu_{\rm min} \approx 70$ Hz for $S_{\rm ns} =  0.005$, in rough agreement with the estimated scaling. For $T_h < 10^{11}$ K and $S_{\rm ns} > 0.05$ we obtain $\nu_{\rm min} \gtrsim 50$ Hz.

\subsubsection{Magnetic Field Effects}

 The frequency at which the magnetic dipole and gravitational radiation spin-down rates are equal is
 \begin{equation}
 \nu_{{\rm kHz}} |C_\alpha|^{2/3} \approx 0.08 M_{1.4}^{-2/3} B_{13}^{2/3}.
 \end{equation}
 Higher $\nu_{\rm kHz}\vert C_\alpha\vert^{2/3}$ implies spindown is mainly via gravitational
radiation rather than magnetic dipole radiation, and vice-versa. In terms of the scaled r-mode amplitude the equation becomes
 \begin{eqnarray}
 \tilde \Omega |\bar{C}_\alpha|^{2/3} &\approx& 1.1 B_{13}^{2/3} M_{1.4}^{-4/3} R_{12.5}^2 \times \\ \nonumber
 && \left(\frac{\kappa}{1.25}\right)^{2/3} \left(\frac{2 \times 10^{-4}}{|\delta \tilde \omega|}\right)^{2/3} \left(\frac{\tilde \omega_\beta \tilde \omega_\gamma}{0.38 \times 0.28}\right)^{1/3},
 \end{eqnarray}
 where $\bar{C}_\alpha(\tilde \Omega)$ is a function of $\tilde \Omega$ with the exact angular velocity dependence changing with the type of evolution.  The magnetic dipole spin-down torque becomes larger than the gravitational driving torque for $|\bar{C}_\alpha| \lesssim 4$ at  $\nu \lesssim 600 $Hz and $B = 10^{13}$ G.  While our simulations reach r-mode amplitudes as high as $|\bar{C}_\alpha| \sim 20$, the amount of time spent at high r-mode amplitudes is low and at $B = 10^{13}$ G the spin-down timescale is dominated by dipole spin-down for most evolutions. An r-mode amplitude of $|\bar{C}_\alpha| \lesssim 4$ corresponds to $C_\alpha  \lesssim 4 \times 10^{-2}$ and a Schenk {\it et al} \cite{Schenk} amplitude of $|c_\alpha| = |C_\alpha|/\sqrt{\Omega} \lesssim 7 \times 10^{-4}$.
 
 Typically, for $B \sim 10^{13}$ G we obtain r-mode instability timescales of a few years. For $B \sim 10^{12}$ G the spin-down timescale grows to $\sim$ 10 years or more. This timescale is also dependent on the frequency at which the spin-down stops, which is partially determined by the shape of the r-mode stability curve.
 
For low magnetic fields $B \le10^{11}$ G the gravitational radiation torque is in general larger than the magnetic dipole torque and the spin-down timescale ranges from tens to thousands of years depending on the strength of the viscosity and of the cooling. Since there are no observed young millisecond pulsars, scenarios with very long spin-down timescales are probably ruled out by observations.
 %is well below the Brunt-$\rm{V\ddot{a}is\ddot{a}l\ddot{a}}$ frequency for typical magnetic fields $B \sim 10^{12} - 10^{13}$ G.  For larger magnetic fields $B \gtrsim 10^{14} G$, the dipole spin-down starts to dominate buoyancy effects $\nu_B \gtrsim \nu_N$. 
  
 The splitting of the frequencies because of the Lorentz force introduced by the magnetic field is small. The corrections for the inertial modes are about three orders of magnitude larger than for the r-mode, but are smaller than the typical detuning of $10^{-4}$ for moderate magnetic fields of $B \sim 10^{12} - 10^{13}$~G and $n\le 15$ (See Appendix C.)  We include dipole spin-down, but ignore these small frequency corrections. 
 
 We also do not include differential rotation, which, if present, could change the modes of the star significantly \cite{Rezolla}.
\section{Stable Evolutions}
\label{stable}
In this section we examine different types of stable evolutions in more detail. We assume a slippage factor $S_{\rm ns} \approx 0.3$, a fraction of the star above the density threshold for direct URCA reactions $f_{\rm dU} = 0.1$ and a critical temperature in the stellar core for proton superfluidity $T_{\rm p} = 5.0 \times 10^9$ K. The critical temperature for hyperon superfluidity $T_{\rm h}$ is treated as a free parameter and varied between $10^9$ and $10^{10}$ K.  We take a coupling coefficients of $\tilde{\kappa} = 1.25$, a fractional detuning $|\delta \tilde \omega| = |\delta \omega|/\Omega = 2.0 \times 10^{-4}$, and the daughter mode frequencies $\tilde \omega_\beta = 0.38$, $\tilde \omega_\gamma = 0.28$ and viscous damping corresponding to the $n=14$, $m=-5$ and $n=15$, $m=3$ modes.
\subsection{Type \cal{I}}
\label{type1}
\begin{figure}
\begin{center}
\leavevmode
\epsfxsize=250pt
\epsfbox{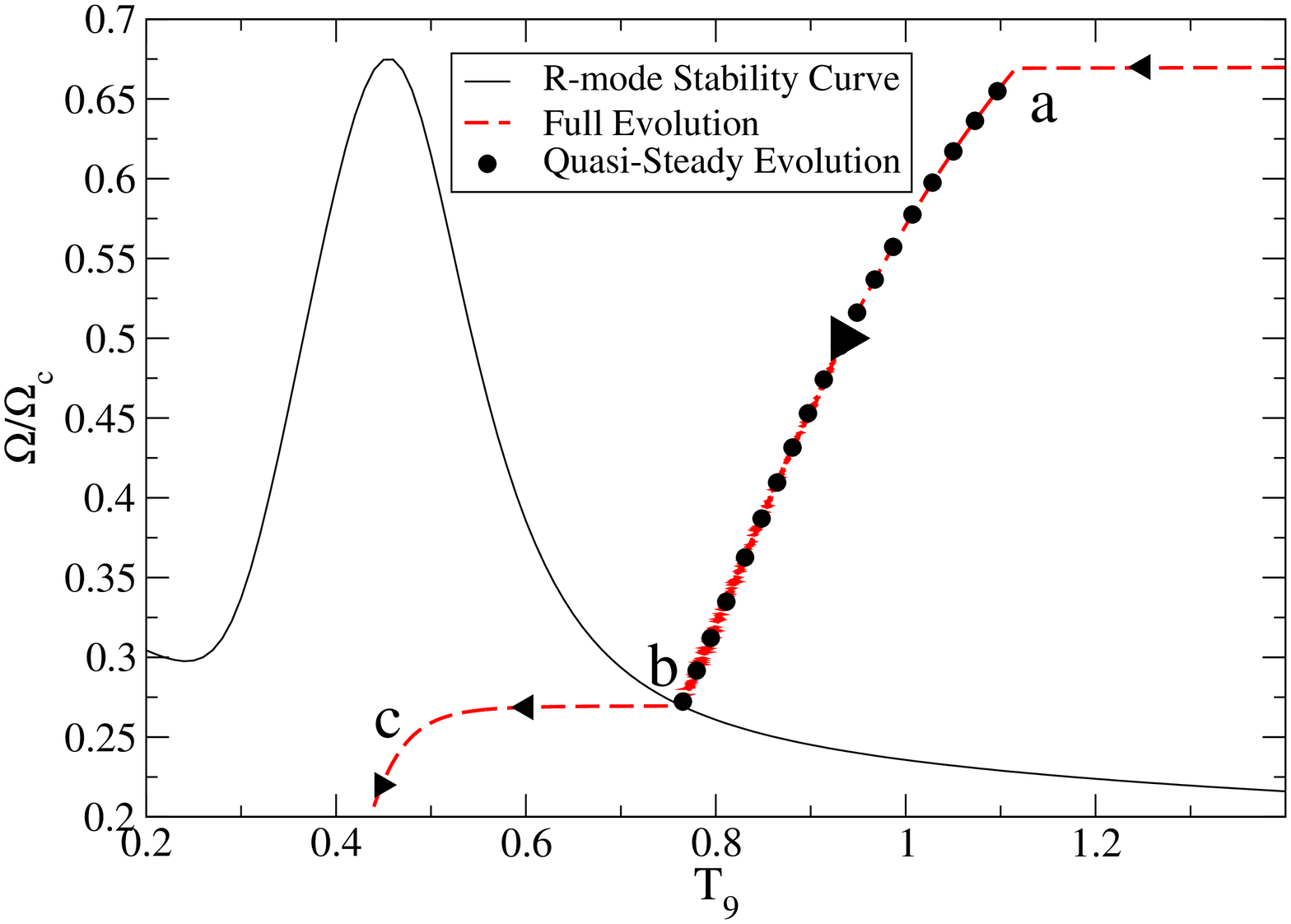}
\newline\newline\newline\newline
\epsfxsize=250pt
\epsfbox{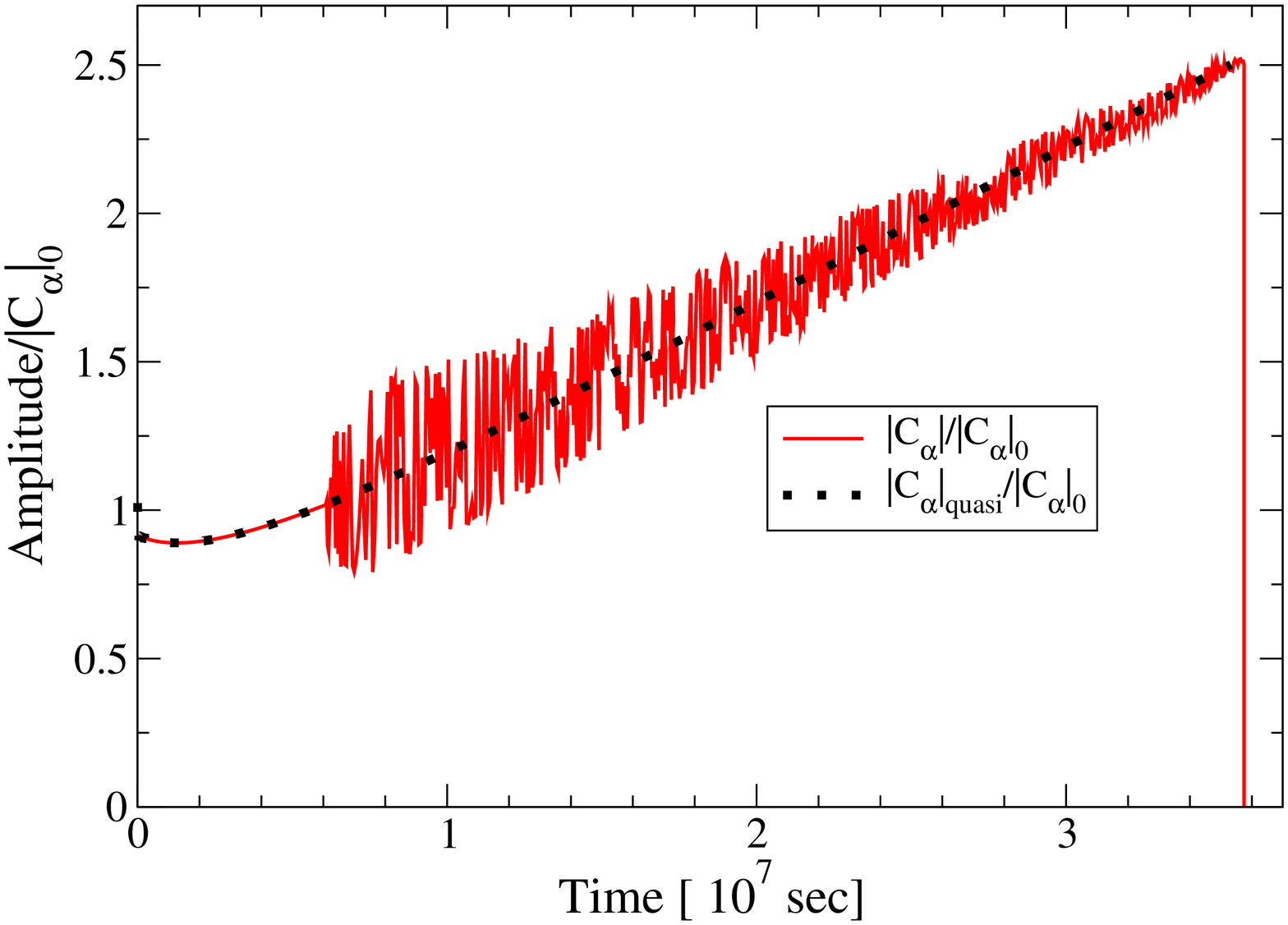}
\caption{ {\bf Type I} evolution. $T_{\rm h} = 2.0 \times 10^9$ K. (a)~The trajectory of the star is shown in the $\tilde\Omega - T_9$ plane. The star cools from $T_{9\,\rm i} = 10$, $\tilde \Omega_{\rm i} = 0.67$ to $T_{9\,\rm a} \approx 1.12$ in about 10 min. At this point the viscous heating is large enough to stop the neutrino cooling. The star then spins down and cools  to $\tilde \Omega_{\rm b} \approx 0.27$, $T_{9\, \rm b} \approx 0.76$ in $t_{\rm a \to b} \approx 1.14$ yr. At point b the star intersects the r-mode stability curve and the r-mode becomes stable. The star starts cooling at constant angular velocity until the cooling rate decreases below the magnetic dipole spin-down rate. (b)~The r-mode and its parametric instability threshold amplitude are shown as a function of time. As before it can be seen that $|\bar{C}_\alpha|_{\rm quasi}$ (Eq.\ \ref{TripletQuasi}) is a good approximation for the average r-mode amplitude until the star crosses in the stable region.} 
 \renewcommand{\arraystretch}{0.75}
 \renewcommand{\topfraction}{0.6}
\label{Tch2}
\end{center}
\end{figure}
 \begin{figure}
\begin{center}
\leavevmode
\epsfxsize=250pt
\epsfbox{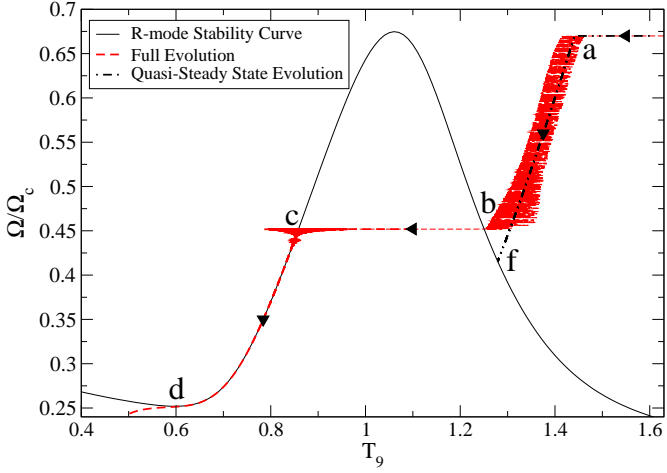}
\caption{ {\bf Mixed I-II } evolution.  $T_{\rm h} = 6.0 \times 10^9$ K. The star cools to $T_{9 \, \rm a} = 1.43$ in about 10 minutes. At this point the dissipative heating of the 3-modes balances the cooling and the star spins down, oscillating around its quasi-steady solution for $t_{\rm a \to b} \approx 0.08$ yr until it intersects the r-mode stability curve at $T_{9 \, \rm b} =1.25$, $\Omega_{\rm b} = 0.45$. It then cools down at constant $\tilde \Omega = \tilde \Omega_{\rm b}$ until the cooling is again balanced by the viscous heating due to the r-mode. The daughter modes do not get excited in this part of the evolution. The star then spins and cools down in thermal equilibrium on the r-mode stability curve for $t_{\rm c \to d} \approx 1.2$ yr until it enters the stable region again at $T_{\rm d} \approx 0.60$, $\tilde \Omega_{\rm d} \approx 0.25$. }
 \renewcommand{\arraystretch}{0.75}
 \renewcommand{\topfraction}{0.6}
\label{A1}
\end{center}
\end{figure}

\begin{figure}
\begin{center}
\leavevmode
\epsfxsize=250pt
\epsfbox{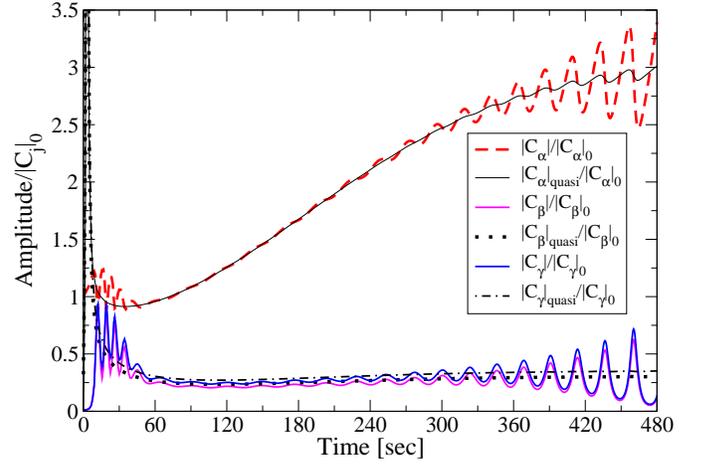}
\newline\newline\newline\newline
\epsfxsize=250pt
\epsfbox{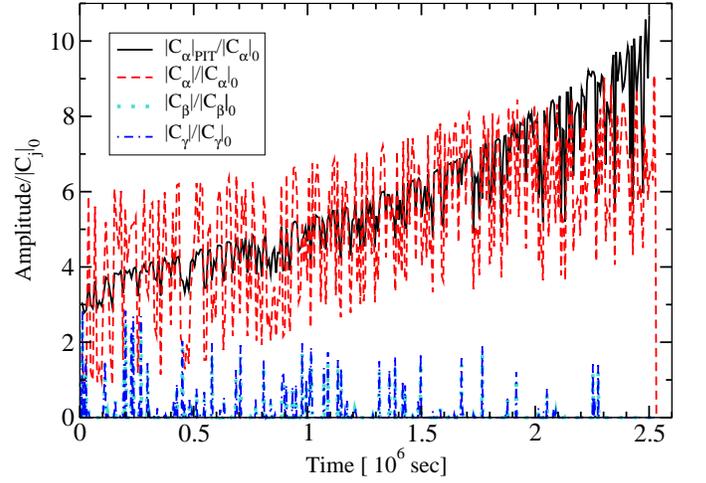}
\newline\newline\newline\newline
\epsfxsize=250pt
\epsfbox{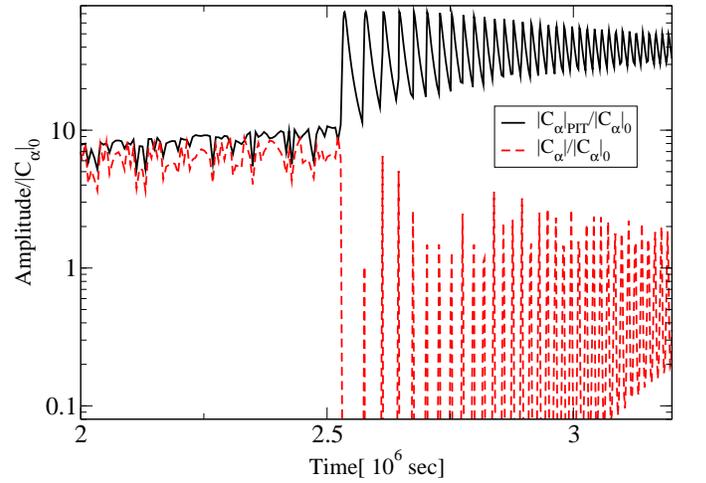}
\caption{{\bf Mixed I-II} evolution. $T_{\rm h} = 6.0 \times 10^9$ K. Mode amplitudes normalized by their zero-viscosity parametric instability threshold shown (a) settling to their quasi-steady states in the first minute or so and then oscillate around these solutions, (b) oscillating in the type I part of the evolution, and (c) switching to a type II oscillation as the star cools from the low $T$ to the high $T$ branch of the stability curve.}
 \renewcommand{\arraystretch}{0.75}
 \renewcommand{\topfraction}{0.6}
\label{A1amps}
\end{center}
\end{figure}

\begin{figure}
\begin{center}
\leavevmode
\epsfxsize=250pt
\epsfbox{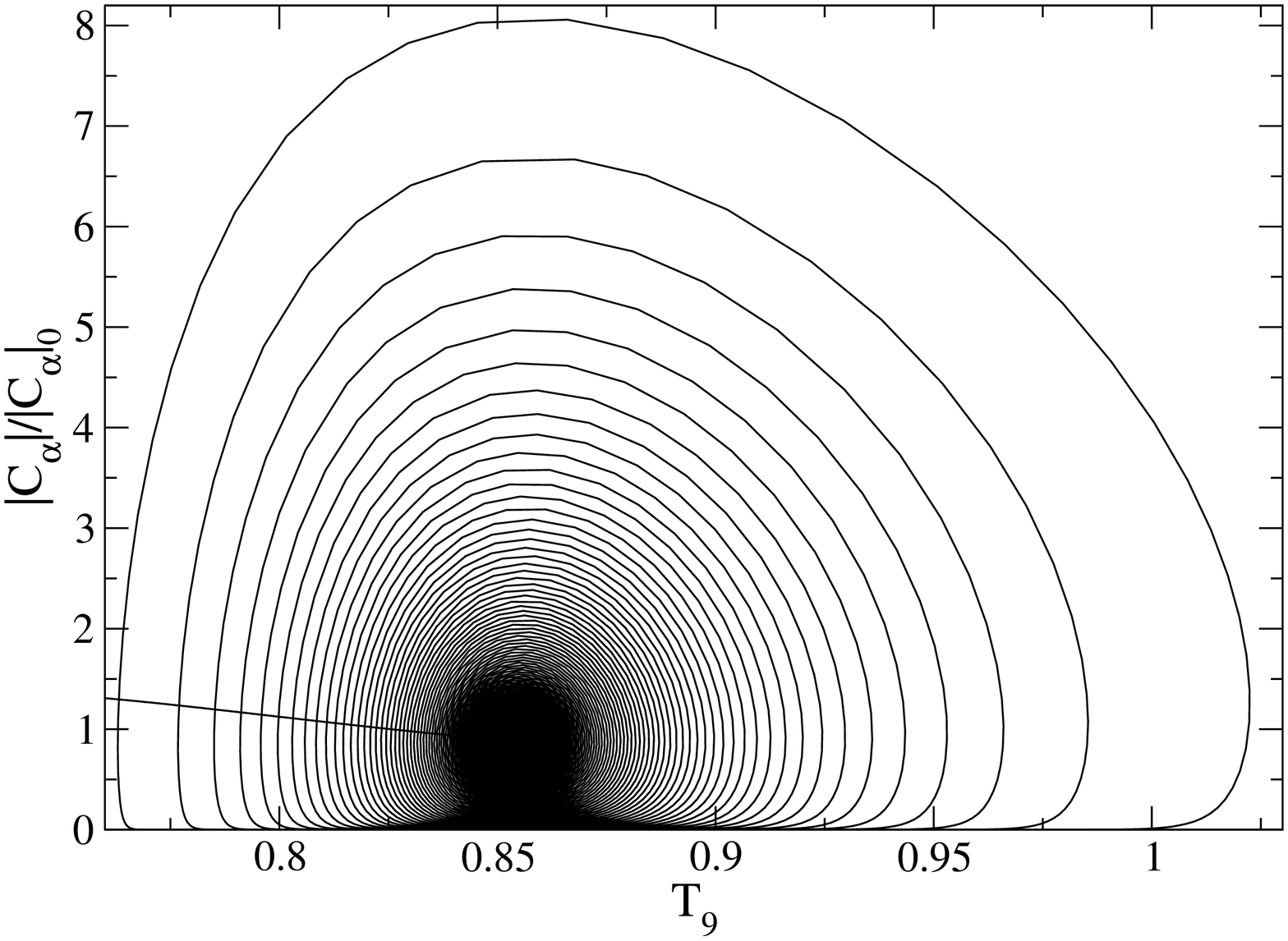}
\newline\newline\newline\newline
\epsfxsize=250pt
\epsfbox{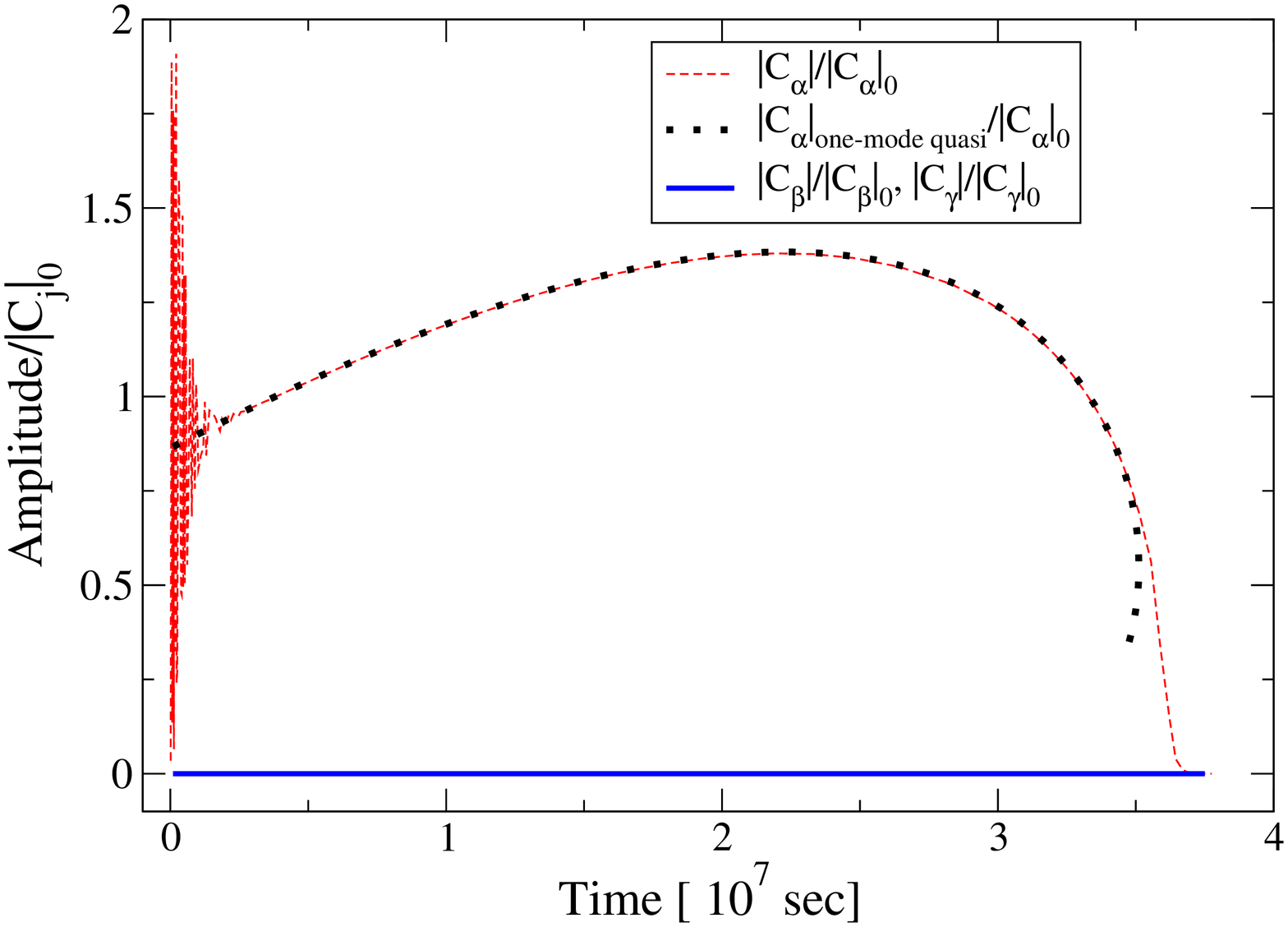}
\caption{ {\bf Mixed I-II} evolution. $T_{\rm h} = 6.0 \times 10^9$ K. The r-mode amplitude for the type II part of the evolution ($T_9 < T_{9 \rm peak}$) is shown as a function of $T_9$ in (a) and as a function of time in (b). Part (a) details the initial oscillation of $|\bar{C}_\alpha|$ around the thermal equilibrium value $|\bar{C}_\alpha|_{\rm C=H\,\, r-mode}$. The r-mode eventually settle to its equilibrium value and continues spinning down along the r-mode stability curve. Part (b) shows the agreement between the r-mode amplitude from the full evolution and the thermal equilibrium r-mode amplitude computed on the r-mode stability curve  $|\bar{C}_\alpha|_{\rm C=H\,\, r-mode} (\tilde\Omega_{\rm CFS}, T_{9\,\rm CFS})$ with $\tilde \Omega$ between $\Omega_{\rm c} = \Omega_{\rm b} \approx 0.45$ and $\Omega_{\rm d} \approx 0.25$.  }
 \renewcommand{\arraystretch}{0.75}
 \renewcommand{\topfraction}{0.6}
\label{A1AmpsRmodeOnly}
\end{center}
\end{figure}
In this subsection we present two evolutions with hyperon superfluidity temperatures of $T_{\rm h} =  2.0 \times 10^9$~K (Fig.\ \ref{Tch2}) and $T_{\rm h} = 6.0 \times 10^9$ K (Fig.~\ref{A1}). The initial angular velocity $\tilde{\Omega}_i = 0.67$ and temperature $T_{9 i} = 10$ are the same in both simulations. The solid arrows in the diagram in Fig.\ \ref{diagram} map the stages that the star goes through for the two trajectories.

The first evolution with $T_{\rm h} =  2.0 \times 10^9$ K  is of type~{\cal I}: the r-mode amplitude passes its parametric instability threshold before reaching the reaching the r-mode stability curve and the star spins down on the $C=H$ curve near the high $T$ side of the r-mode stability curve. Fig.\ \ref{Tch2}(a) shows the trajectory in the $\Omega-T$ plane. The evolution starts in the unstable regime where $\gamma_{GR} >> \gamma_{\alpha \, v}$. The r-mode amplitude increases exponentially until it passes the first parametric instability threshold and excites two inertial modes.  After a brief precursor of about $15$ min the three modes settle in their quasi-stationary states. The star continues to cool at constant angular velocity until it reaches thermal equilibrium at $T_9 \approx 1.12$. This occurs far from the r-mode stability curve $T_{9\, {\rm CFS}} (\tilde \Omega = 0.67) = 0.47$.  The neutron star then spins down and cools slowly oscillating around the $C=H$ curve determined by the quasi-stationary mode amplitudes. Fig.\ \ref{Tch2}(b) shows the r-mode amplitude oscillating around its quasi-stationary solution.  The star intersects the r-mode stability curve at $\tilde \Omega_{\rm b} = 0.27$ and $T_{9\, {\rm b}} = 0.76$. The spin-down timescale $t_{\rm a \to \rm b} \approx 1.1$~yrs at $B = 10^{13}$ G. This timescale is independent of detuning, but changes with $B$.  For $B=10^{11}$ G, $t_{a\to b} \approx 14.3$ yr. Lowering the magnetic field further leaves the timescale unchanged because beyond this $B$ the gravitational radiation torque is larger than the magnetic dipole torque. Point $b$ can be determined analytically by finding the intersection between the r-mode stability curve and the $C=H$ curve. Once the star crosses in the stable region, the amplitudes of the three modes damp to zero. The neutron star cools via neutrino emission at constant angular velocity until the cooling rate becomes slower than the magnetic dipole spin-down rate and then spins down emitting magnetic dipole radiation. 

The second scenario has $T_h = 6 \times 10^9$ K and is a mixed I-II evolution: the star first reaches thermal evolution near the high $T$ side of the r-mode stability curve (type I evolution) and later, after it intersects the stability curve, the star cools to the low $T$ branch of the curve and continues the spin down on this branch (type II evolution). The trajectory in the $\Omega-T$ space is displayed in Fig.~\ref{A1}. The star cools from $T_{9 \rm i} =10$ to $T_9 \approx 2$ in the first minute or so. The r-mode and daughter mode amplitudes settle in their quasi-stationary states at $T_9 \approx 2$ and subsequently oscillate around these solutions (Fig.\ \ref{A1amps}(a)). Increasing $T_h$ shifts the r-mode stability curve to higher $T$ and the cooling is stopped at a higher temperature than in the $T_h= 2\times 10^9$ K case.  At $T_{9 \rm a} \approx 1.43$ the star reaches thermal equilibrium and starts spinning down oscillating around the quasi-steady $C=H$ curve for about $t_{\rm a\to b}\approx 0.1$ yr. Eventually, the oscillations become unstable and their amplitude increases until the star intersects the r-mode stability curve at $(T_{9 \,\rm b}, \tilde \Omega_{\rm b}) = (1.25, 0.45)$. By this point the neutron star has lost approximately 32\% of its initial angular velocity, while the temperature decreased by about 10\% after the star first reached thermal equilibrium. All points on the high $T$ branch of the r-mode stability curve correspond to unstable thermal equilibrium (see Appendix D). So, when the trajectory reaches this branch of the stability curve, the star cools at  constant $\Omega$ through the stable region until it intersects the r-mode stability curve again (assuming the cooling is faster than the magnetic spin-down rate). For this evolution the cooling through the stable region takes about $t_{b \to c} = 14$ hours.

For the r-mode amplitude, the transition between the two unstable regions on the different sides of the
stability curve is shown in Fig.\ \ref{A1amps}(c). As the star cools through the stable region the r-mode
amplitude decreases and the parametric instability threshold rises. Once the r-mode becomes unstable again the star is trapped in thermal equilibrium on the r-mode stability curve. The evolution is now of type II. The settling oscillations of the r-mode amplitude are shown in Fig.\ \ref{A1AmpsRmodeOnly}(a). The stability curve acts as an attractor. Fig.\ \ref{A1AmpsRmodeOnly}(b) plots the r-mode amplitude together with its quasi-steady one mode solution as a function of time. The agreement is very good until the star intersects the stability curve again. At this point the r-mode amplitude damps to zero and the star continues to spin-down due to the magnetic dipole torque.

\subsection{Type \cal{II}}
\label{type2}
\begin{figure}
\begin{center}
\leavevmode
\epsfxsize=250pt
\epsfbox{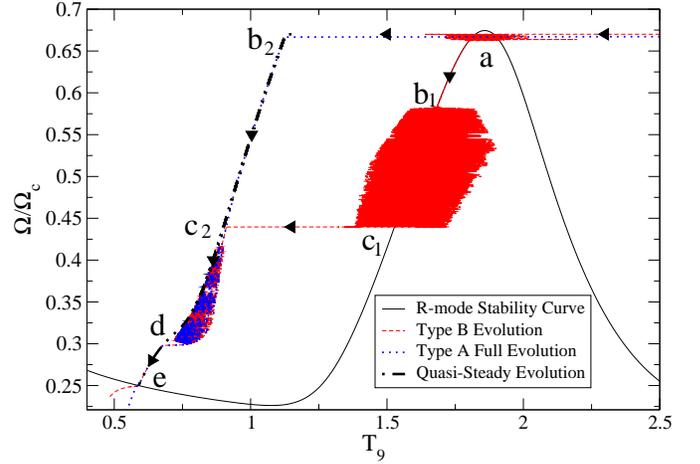}
\caption{The trajectory of a neutron star in the $\tilde \Omega - T_9$ plane is shown for a (1)  {\bf mixed II-III} and (2) {\bf type III} evolution with $T_h = 1.2 \times 10^{10}$ K. In both cases star cools to $T_{9 \, \rm a} = 1.82$ in about 6 minutes and crosses in the unstable region bounded by the low T branch of the r-mode stability curve. In evolution 1 the cooling is balanced by dissipative heating close to the low T branch of the stability curve. The star settles on the stability curve and spins down on it, $t_{\rm a \to c_1} = 6.5 \times 10^5$ sec. As the temperature and spin frequency decrease the thermal equilibrium becomes unstable leading to growing thermal oscillations. The r-mode passes its parametric instability threshold again and the star eventually cools ($t_{c_1 \to c_2} = 1.2 \times 10^4$ sec) to the next $C=H$ curve turning into a type III evolution. In  evolution 2 the star cools at constant  $\tilde \Omega$ for $\sim 2$ hours until $T_{9\rm b_2} \approx 1.13$. It then spins down for $t_{\rm b_2 \to e} \approx 1.4$ yr on a $C=H$ curve determined by the quasi-steady states of all three modes until the trajectory intersects the r-mode stability curve again. }
 \renewcommand{\arraystretch}{0.75}
 \renewcommand{\topfraction}{0.6}
\label{Tch12}
\end{center}
\end{figure}

\begin{figure}
\begin{center}
\leavevmode
\epsfxsize=250pt
\epsfbox{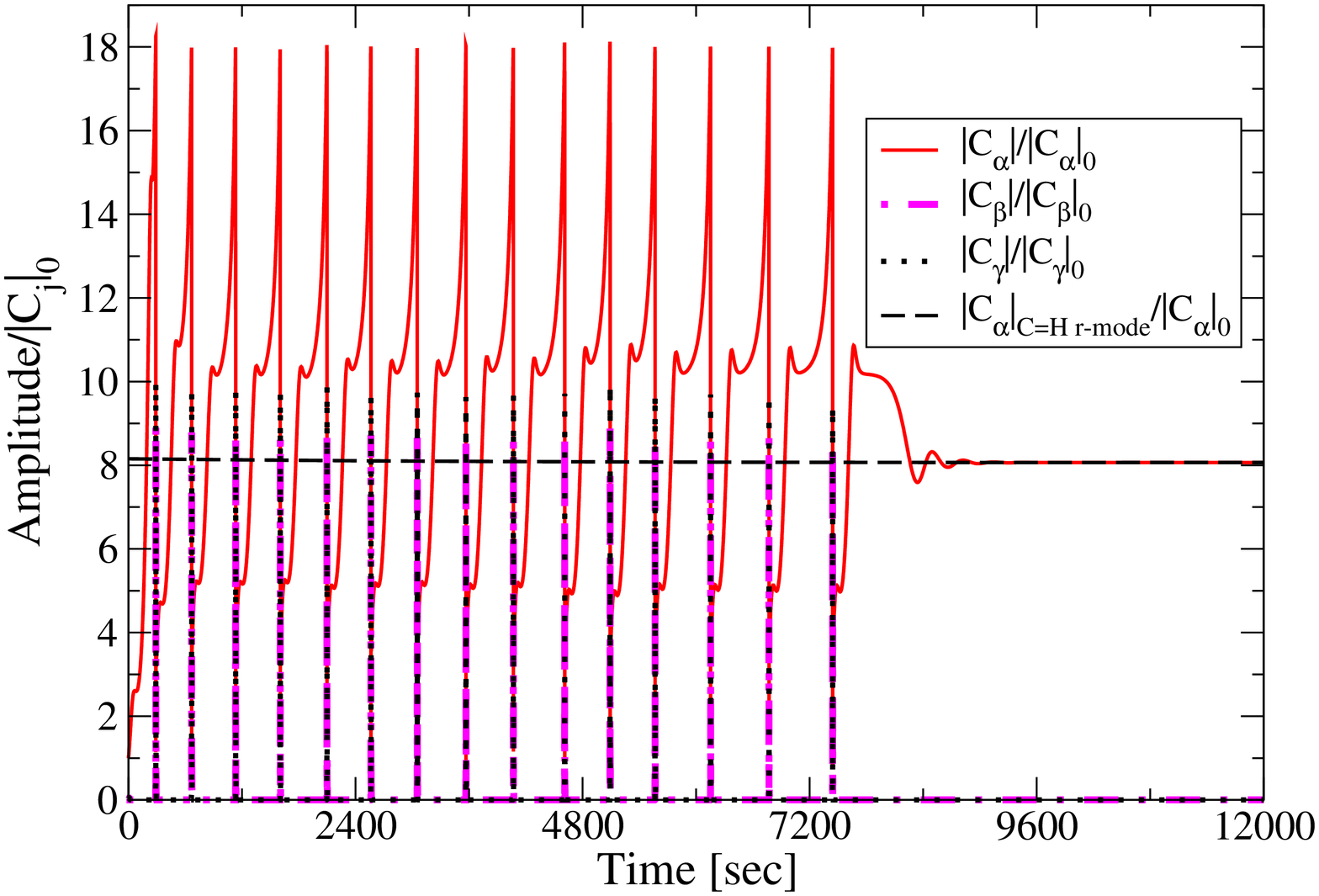}
\newline\newline\newline\newline
\epsfxsize=250pt
\epsfbox{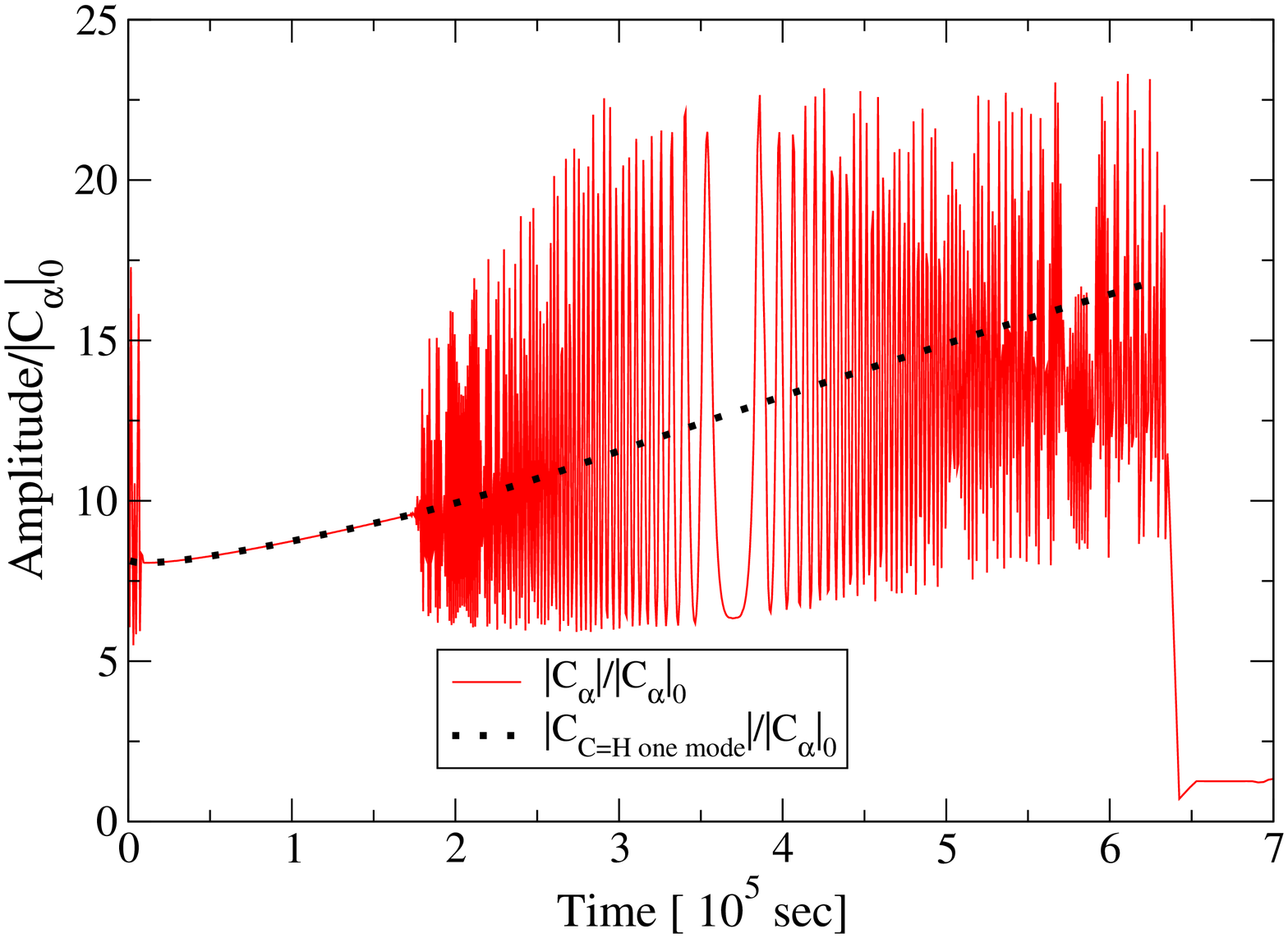}
\caption{ {\bf Mixed II-III} evolution. $T_h = 1.2 \times 10^{10}$ K.  (a) The initial r-mode $|\bar C_\alpha|$ and daughter mode $|\bar C_\beta|$, $|\bar C_\gamma|$ amplitudes. When the evolution reaches thermal equilibrium on the r-mode stability curve, the r-mode settles to its one-mode equilibrium value $|\bar C_\alpha|_{\rm C=H \, one-mode}$ and the daughter mode oscillations damp to zero. (b) The r-mode amplitude and its thermal equilibrium solution $|\bar C_\alpha|_{\rm C=H \, one-mode}$ are shown for the type II part the evolution.   $|\bar C_\alpha|_{\rm C=H \, one-mode}$ is a good approximation for the average r-mode amplitude. }
 \renewcommand{\arraystretch}{0.75}
 \renewcommand{\topfraction}{0.6}
\label{AmpTch12}
\end{center}
\end{figure}

The different stages of type II evolutions are mapped by the dashed-lined arrows in Fig.\ \ref{diagram}.
In this scenario the star reaches the r-mode stability curve, cools through the stable region, and settles in thermal equilibrium along the low $T$ branch of the r-mode stability curve. These evolutions can be approximated by a one-mode model.

 For the $T_{\rm h} = 6.0 \times 10^9$ K and $\tilde{\Omega}_i = 0.60$, type II evolutions occur when 
 $\bar{C}_\alpha(0) < 3.0 \times 10^{-4}$. They last about 1.2 yr for $B = 10^{13}$ G. This timescale increases for lower B and is independent of the detuning. In these evolutions the r-mode amplitude does not grow above its parametric instability threshold. When $\bar{C}_\alpha(0) \ge 3.0 \times 10^{-4}$ mixed I-II evolutions occur. These latter evolutions are similar to the one in Fig.\ \ref{A1} and follow the same $C=H$ curve.
 
We next consider $T_h = 1.2 \times 10^{10}$ K. At $\tilde{\Omega}_i = 0.67$, initial
r-mode amplitudes $|\bar{C}_\alpha|(0) > 5.0 \times 10^{-5}$ lead to mixed evolutions that
start as type II and end as type III. $|\bar{C}_\alpha|(0) \le 5.0 \times 10^{-5}$ result in type III evolutions
and will be described in the next subsection. The trajectories for these two evolutions are presented in Fig.\ \ref{Tch12}.

 The transition r-mode amplitude between different types of evolutions $|\bar{C}_\alpha|(0)$ is very sensitive to both $T_h$ and $\tilde \Omega_i$. Lowering $\Omega_i$ decreases the driving rate of the r-mode. Similarly, increasing $T_h$ shifts the r-mode stability curve leading to higher viscosity at higher temperatures and hence lower growth rate for the r-mode amplitude. Lowering the growth rate of the instability makes it more likely that the r-mode amplitude will reach its parametric instability threshold later. This increases the value of the transition amplitude favoring cooling through the first $C=H$ curve.

In mixed II-III evolutions the r-mode amplitude passes its first parametric instability threshold close to the low T branch of the r-mode stability curve. The dissipative heating from the three modes stops the cooling close to the low $T$ branch of the r-mode stability curve. Fig.\ \ref{AmpTch12}(a) plots the initial evolution of the three mode amplitudes. As the trajectory of the star in the $\Omega-T$ plane converges toward the r-mode stability curve, the r-mode amplitude settles into its one mode quasi-steady solution and the daughter modes damp to zero.  Fig.\ \ref{AmpTch12}(b) shows the r-mode amplitude oscillating around its quasi-stationary state. When the equilibrium becomes unstable and 
the oscillations grow, the r-mode passes its parametric instability threshold again. The star escapes this equilibrium and cools to the next $C=H$ curve in about 4 hours. It then spins down on this curve until the trajectory intersects the r-mode stability curve again. Afterwards the mode amplitudes  damp to zero leaving the magnetic dipole torque to continue to spin down the star.
\subsection{Type \cal{III}}
\label{type3}
\begin{figure}
\begin{center}
\leavevmode
\epsfxsize=250pt
\epsfbox{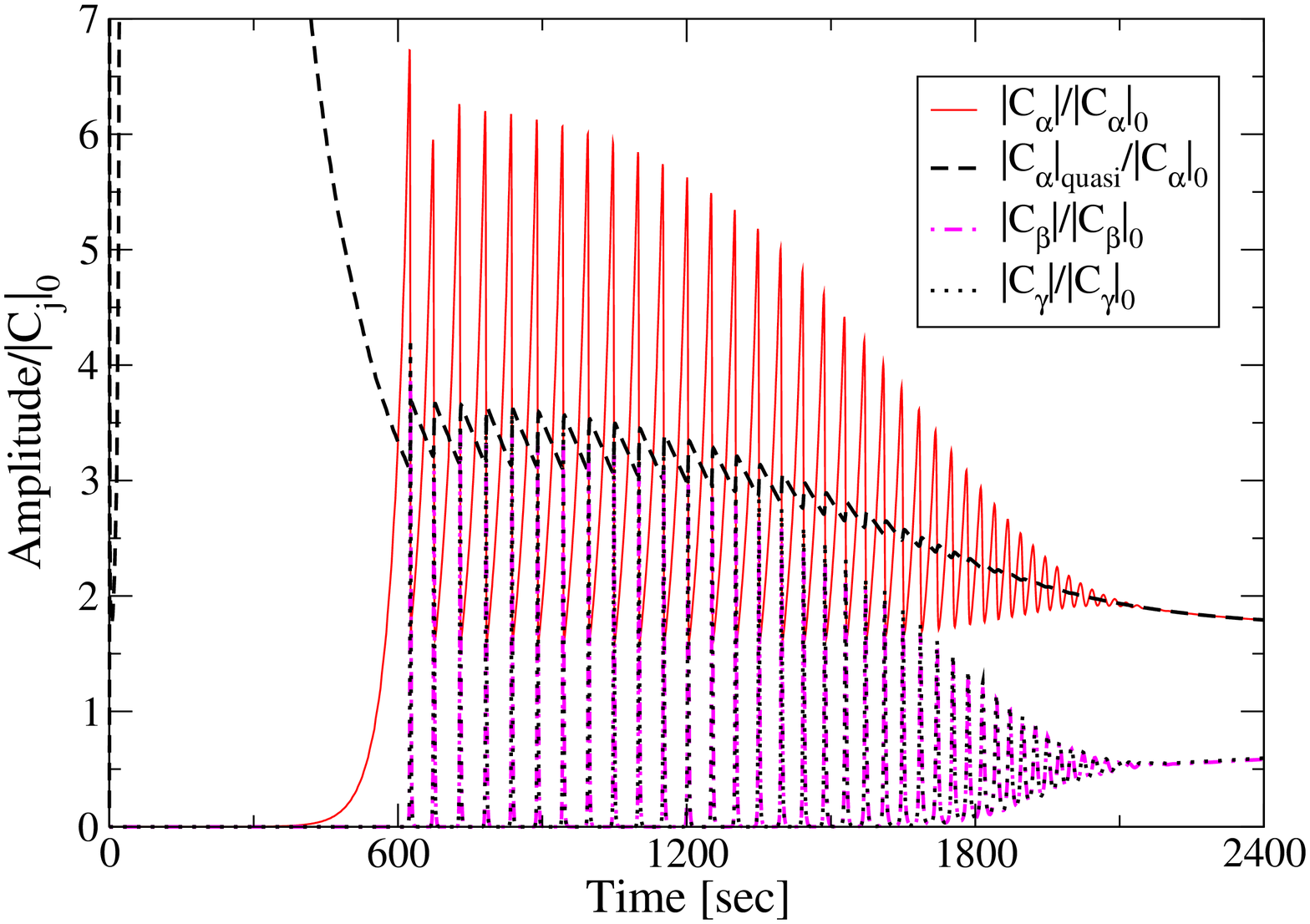}
\epsfxsize=250pt
\newline\newline\newline\newline
\epsfbox{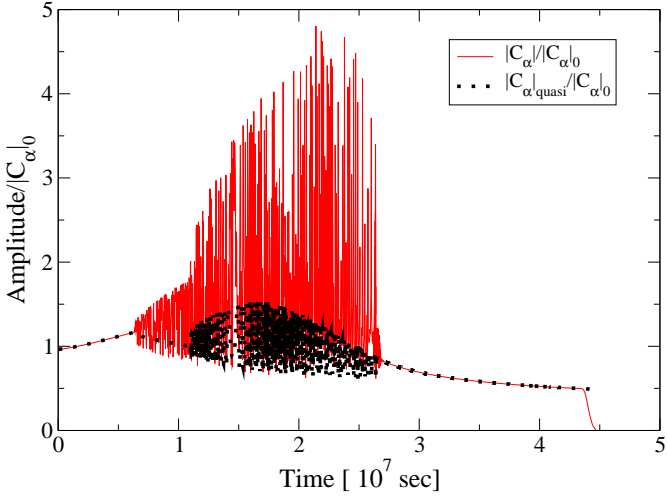}
\caption{{\bf Type III} evolution. $T_h = 1.2 \times 10^{10}$ K. (a) The amplitudes of the r-mode $|\bar C_\alpha|$ and of the two daughter modes $|\bar C_\beta|$, $|\bar C_\gamma|$ are shown settling to their quasi-steady states for the first 40 minutes. (b)~The r-mode amplitude and its lowest parametric instability threshold is shown as a function of time for the whole evolution. The quasi-steady solution, which coincides with the parametric instability threshold, is a good approximation for the average r-mode amplitude with an almost exact agreement in the non-oscillatory part of the trajectory.}
 \renewcommand{\arraystretch}{0.75}
 \renewcommand{\topfraction}{0.6}
\label{amplitudesLHS}
%placeholder
\end{center}
\end{figure}

Type \cal{III} evolutions occur when the star cools through the stable region before the r-mode amplitude reaches its parametric instability threshold, overshoots the thermal equilibrium on the low $T$ branch of the stability curve and then settles on the next $C=H$ curve.  As an example we are examining evolution $2$ in Fig.\ \ref{Tch12}. This simulation has $T_{\rm h} = 1.2 \times 10^{10}$ K, and the usual $\Omega_{\rm i} = 0.67$, $T_{9\, \rm i} = 10$, $B = 10^{13}$ G. Any $|\bar{C}_\alpha|(0) \le 5.0 \times 10^{-5}$ for this value of  $T_h$ leads to approximately the same type III evolution.  The transition amplitude $|\bar{C}_\alpha|(0)$ changes with $\tilde \Omega_i$ and $T_h$ as discussed before. Lower $\tilde{\Omega}_i$ (higher $T_h$) lead to higher transition amplitudes. %because they occur in the thermal equilibrium region at the lowest temperature.

Fig.\ \ref{amplitudesLHS}(a) shows the initial r-mode amplitude and daughter mode amplitudes settling into their quasi-stationary states in the first $\sim 40$ minutes of the evolution. The star continues cooling and reaches thermal equilibrium at $t_{\rm b_2} \approx 6.5 \times 10^5$ sec.  It briefly oscillates and then settles on the quasi-steady $C=H$ curve. Fig.\ \ref{amplitudesLHS}(b) displays the
r-mode amplitude $|\bar{C}_\alpha|$ and its quasi-stationary solution, which coincides with the parametric instability threshold amplitude. The agreement between the full evolution $|\bar{C}_\alpha|$ and its quasi-steady counterpart is very good for the first 0.2 yrs or so. As the star cools and spins down slowly the viscosity decreases and the star develops thermal oscillations that are initially unstable with growing amplitude. The thermal oscillations become stable again when boundary layer viscosity dominates bulk viscosity and the temperature dependence of the viscosity changes. For a detuning of $\delta \omega/2 \Omega = 10^{-4}$ and $B = 10^{13}$ G, the spin-down lasts about 1.4 yrs. Afterwards, trajectory of the star intersects the r-mode stability curve again.
This spin-down timescale is dependent on the size of the detuning and the strength of the magnetic 
field.  For $B \le 10^{11}$ G ($t_{b_2\to e} \approx 130$) the gravitational radiation torque dominates the spin-down. When $\delta \omega \to 0$ this type of evolution no longer exists.  The average detuning is roughly inversely proportional to the number of direct couplings to the r-mode (see Sec.\ \ref{validity}). So, as the model becomes more sophisticated and more modes are added due to, for example, superfluidity effects, the average detuning may become several orders of magnitude smaller and this scenario could disappear.  

%the neutrino cooling decreases with $T_9$ and the star heats until thermal balance is reached again.  

\section{Runaway Evolutions}
\label{unstable}
\begin{figure}
\begin{center}
\leavevmode
%\epsfxsize=250pt
%\epsfbox{Fig11b.eps}
\epsfxsize=250pt
\epsfbox{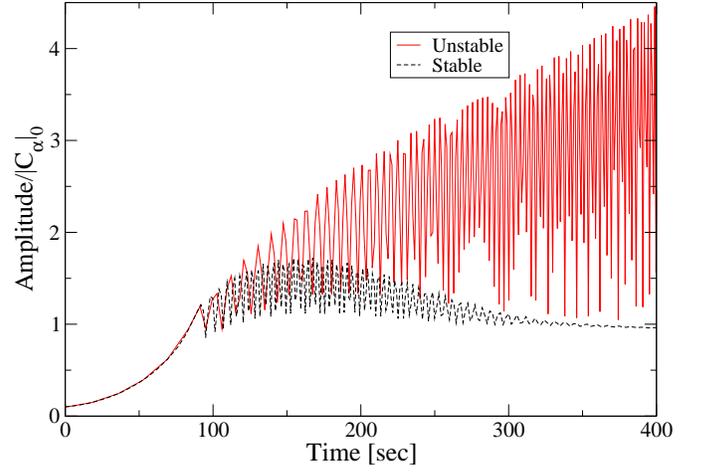}
\caption{{\bf Runaway evolutions.} $T_{\rm h} = 6 \times 10^9$ K.  The r-mode amplitude
 is shown versus time for the two evolutions with different viscosity:  for the stable case ( $A_{\rm hb} = 2.0 \times 10^2$ sec$^{-2}$) $|\bar C_\alpha|$ oscillates and settles close to its quasi-steady state while in the unstable case ($A_{\rm hb} = 1.0 \times 10^2$ sec$^{-2}$) it continues to grow. Both evolutions have $\bar{C}_\alpha(0) = 0.1$.}
 \renewcommand{\arraystretch}{0.75}
 \renewcommand{\topfraction}{0.6}
\label{StableUnstable}
\end{center}
\end{figure}

\begin{figure}
\begin{center}
\leavevmode
\epsfxsize=250pt
\epsfbox{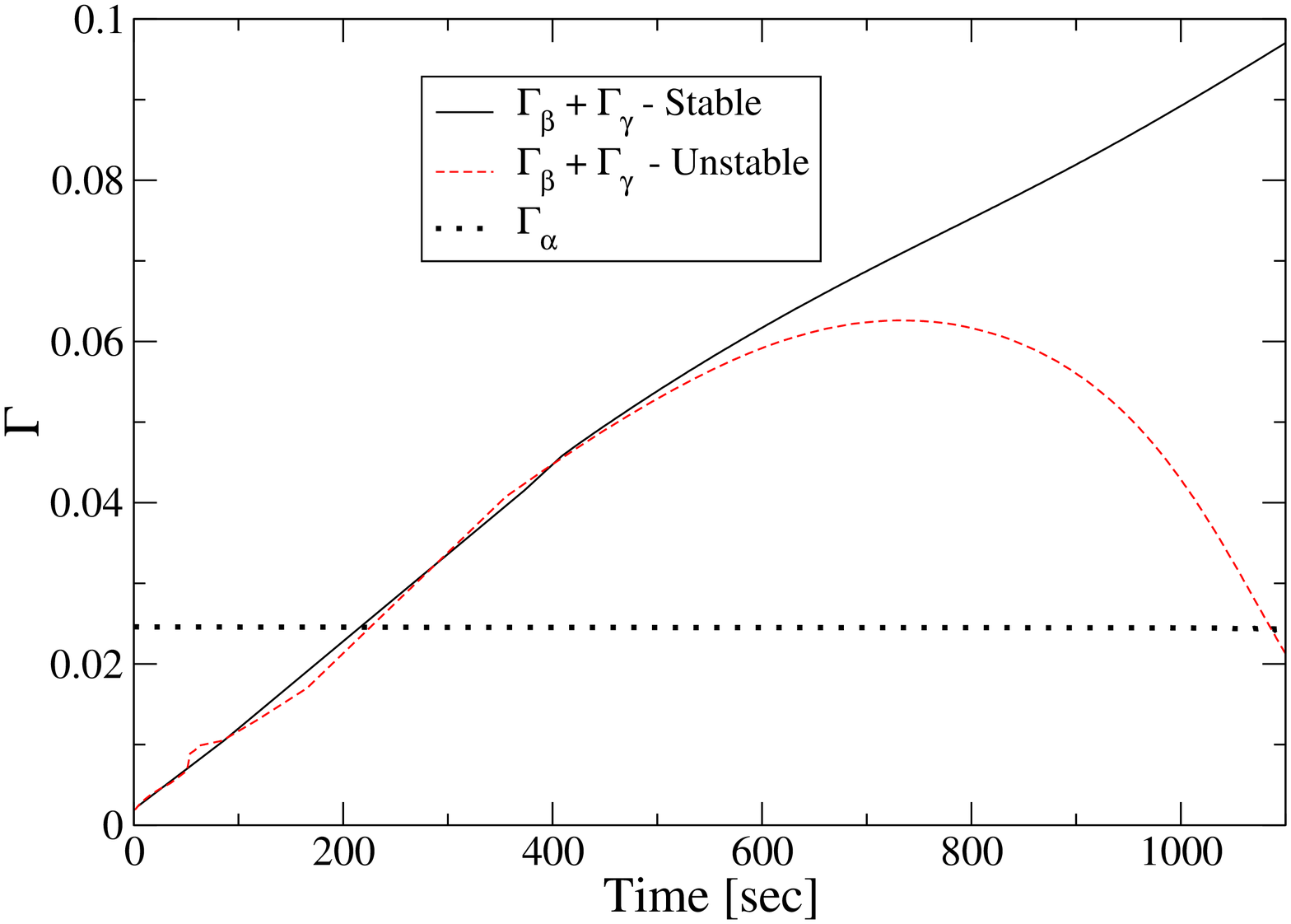}
\newline\newline\newline\newline
\epsfxsize=250pt
\epsfbox{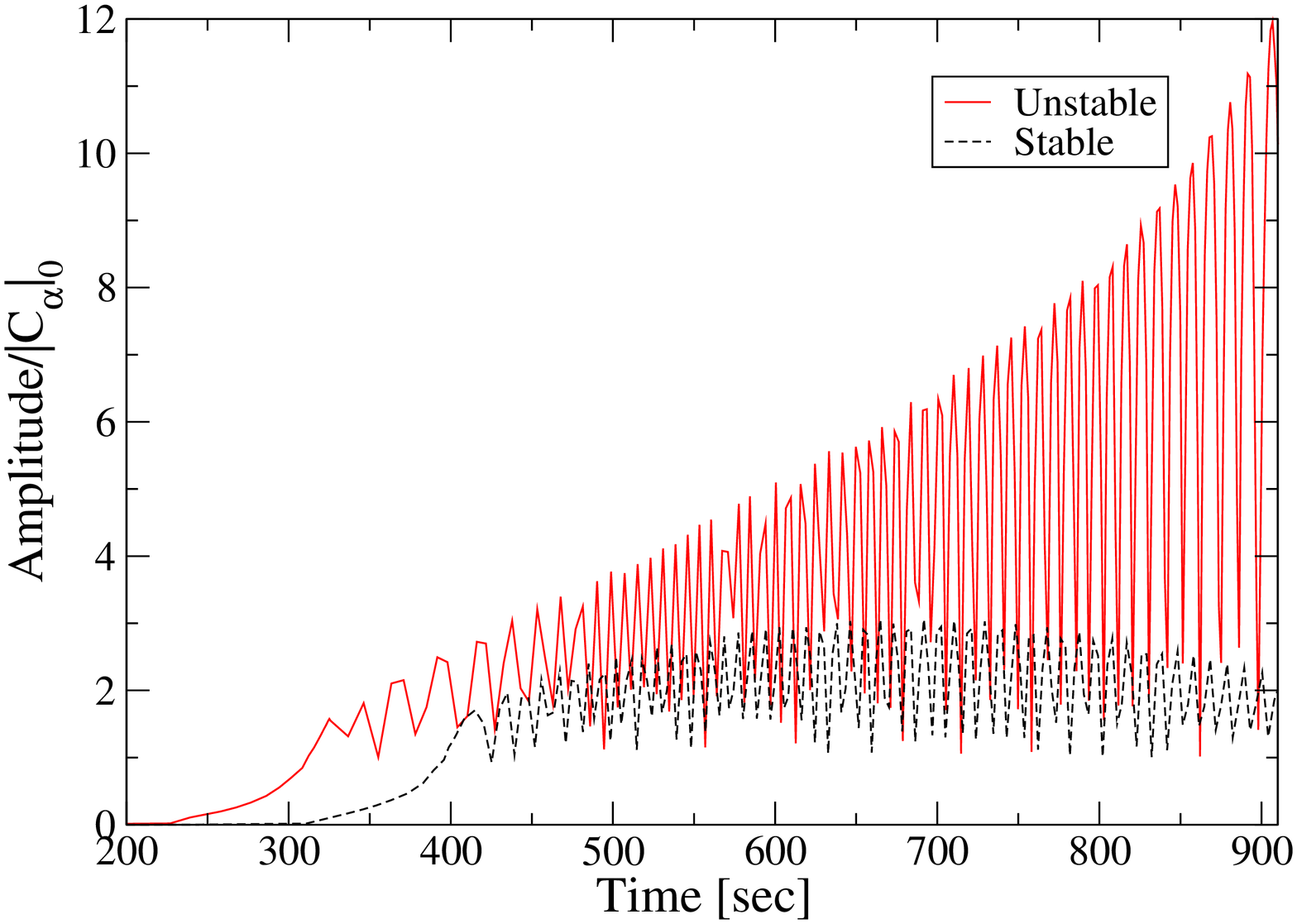}
\caption{{\bf Runaway evolutions.} $T_{\rm h} = 6 \times 10^9$ K. A similar runaway occurs for low cooling. Here $f_{\rm dU} = 0.001$, $A_{\rm hb} = 5.0\times 10^3$ sec$^{-2}$ for both evolutions. The initial amplitudes are
$|\bar C_\alpha|(0)_{\rm stable} = 2.0 \times 10^{-5}$ and $|\bar C_\alpha|(0)_{\rm unstable} = 2.0 \times 10^{-4}$. }
 \renewcommand{\arraystretch}{0.75}
\label{StableUnstable2}
\end{center}
\end{figure}

Mathematically, one can easily see that  $\Gamma_\alpha > \Gamma_\beta + \Gamma_\gamma$ corresponds to an unstable evolution by looking at the second derivative of the r-mode amplitude evolution equation:
\begin{eqnarray}
\frac{d^2 \bar{C}_\alpha}{d\tilde \tau^2} &=& \frac{d\bar{C}_\alpha}{d\tilde \tau} (\Gamma_\alpha - \Gamma_\beta - \Gamma_\gamma) +  \\ \nonumber
&& \bar{C}_\alpha \left[\Gamma_\alpha (\Gamma_\beta + \Gamma_\gamma) - \frac{|\bar C_\beta|^2 + |\bar C_\gamma|^2}{4 \tilde \Omega}\right],
\end{eqnarray}
where $\Gamma_\alpha = \tilde{\gamma}_\alpha/(\tilde \Omega |\delta \tilde{\omega}|) = \gamma_\alpha/\delta \omega$, $\Gamma_\beta = \tilde{\gamma}_\beta/(\tilde \Omega |\delta \tilde{\omega}|) = \gamma_\beta/\delta \omega$, and $\Gamma_\gamma = \tilde{\gamma}_\gamma/(\tilde \Omega |\delta \tilde{\omega}|) = \gamma_\gamma/\delta \omega$.
If $\Gamma_\alpha > \Gamma_\beta + \Gamma_\gamma$ then the r-mode is unstable no matter how large the daughter modes become.

Assuming a solution of the r-mode amplitude of the form $\bar{C}_\alpha \propto \exp{(st)}$ and taking the daughter modes to be constant gives the roots
\begin{eqnarray}
s_{\pm} &=& \frac{1}{2} (\Gamma_\alpha - \Gamma_\beta - \Gamma_\gamma)  \pm  \frac{1}{2}.\left\{(\Gamma_\alpha - \Gamma_\beta - \Gamma_\gamma)^2 \right.  \\ \nonumber 
&& \left. + 4\left [\Gamma_\alpha (\Gamma_\beta + \Gamma_\gamma) - \frac{|\bar{C}_\beta|^2 + |\bar{C}_\gamma|^2}{4 \tilde \Omega}\right]\right\}^{1/2} \\ \nonumber
&=& \frac{1}{2} \left\{ (\Gamma_\alpha - \Gamma_\beta - \Gamma_\gamma)  \right. \\ \nonumber
&& \left. \pm \sqrt{(\Gamma_\alpha + \Gamma_\beta + \Gamma_\gamma)^2 - \frac{|\bar{C}_\beta|^2 + |\bar{C}_\gamma|^2}{\tilde \Omega}}
\right\}
\end{eqnarray}
If $\Gamma_\alpha < \Gamma_\beta + \Gamma_\gamma$ then the $s_{-}$ mode is always stable and
the $s_{+}$ mode is stable only if
\begin{equation}
|\bar{C}_\beta|^2 + |\bar{C}_\gamma|^2 < 4 \Gamma_\alpha (\Gamma_\beta + \Gamma_\gamma) \tilde \Omega.
\end{equation}
In Fig.\ \ref{StableUnstable} we compare two evolutions: one with lower viscosity ($A_{\rm hb} = 1.0 \times 10^2$ sec$^{-2}$) that is unstable and the other with slightly higher hyperon bulk viscosity ($A_{\rm hb} = 2.0 \times 10^2$ sec$^{-2}$) that is stable. In the unstable evolution, $\Gamma_\beta + \Gamma_\gamma$ decrease below $\Gamma_\alpha \approx \gamma_{GR}/\delta \omega$. The amplitude oscillations in the stable evolution damp and settle close to the r-mode quasi-stationary solution, while, in the unstable evolution,  the r-mode amplitude continues to grow passing several parametric instability thresholds and exciting more inertial modes.

In our model, for $T > T_{\rm peak}$ the hyperon bulk viscosity is a decreasing function of temperature. Another way to make the r-mode amplitude more likely to overshoot its first parametric instability threshold is to lower the neutrino cooling. Slower cooling causes the star to spend more time at high temperatures and low viscosity. For  $f_{\rm dU} = 0.001$, evolutions are unstable for $|\bar C_\alpha|(0) \ge 2 \times 10^{-4}$ (See Fig.\ \ref{StableUnstable2} for an example.) As before, when the evolution is unstable, $\Gamma_\beta + \Gamma_\gamma$ decreases below $\Gamma_\alpha$ and the three modes cannot stop the growth of the instability.  

%\caption{(a) The growth rate of the r-mode $\tilde{\gamma}_\alpha = \tilde{\gamma}_{GR} - \tilde{\gamma}_{\alpha \, v}$ and the sum of the viscous damping rates for the two daughter modes $\tilde{\gamma}_\beta + \tilde{\gamma}_\gamma$ are shown versus time for a different scenario. (b) The r-mode amplitude $\bar C_\alpha$ is shown versus time for the two evolutions.%Here $T_{\rm c \, h} = 1.2 \times 10^{10}$ K, $f_{\rm dU} = 0.0$, $A_{\rm hb} = 5.0\times 10^3$}
%%The initial amplitudes differ with $|\bar C_\alpha|(0)_{\rm stable} = 6.0 \times 10^{-5}$ and $|\bar C_\alpha|(0)_{\rm unstable} = 7.0 \times 10^{-5}$.
%%placeholder}
\section{Detection of Gravitational Waves}
\label{detection}

%\begin{figure}
%\begin{center}
%\leavevmode
%\epsfxsize=250pt
%\epsfbox{Figures/href1.eps}
%\epsfxsize=250pt
%\epsfbox{Figures/hd2.eps}
%\caption{The gravitational wave amplitude $h_0$ multiplied by the distance to the source is shown as  a function of time for the evolutions discussed in this paper: (a) Evolution A, D and the $T_{\rm c \, h} = 2.0 \times 10^9$ K evolution presented in Sec.\ \ref{stable} (b) The part of Evolution C for which the trajectory oscillates around the r-mode stability curve. The second part of the evolution overlaps with D. 
%The Advanced LIGO and Einstein Telescope (ET) lines are computed multiplying the gravitational wave amplitude from Eq.\ (xx) by 10 Mpc. We are using a nominal broad band configuration for Advanced LIGO and a proposed curve for the Einstein Telescope from Ref.\ \cite{Watts}.}
%%The initial amplitudes differ with $|\bar C_\alpha|(0)_{\rm stable} = 6.0 \times 10^{-5}$ and $|\bar C_\alpha|(0)_{\rm unstable} = 7.0 \times 10^{-5}$.
%%placeholder
% \renewcommand{\arraystretch}{0.75}
% \renewcommand{\topfraction}{0.6}
%\label{hdplot}
%\end{center}
%\end{figure}
\begin{figure}
\begin{center}
\leavevmode
\epsfxsize=250pt
\epsfbox{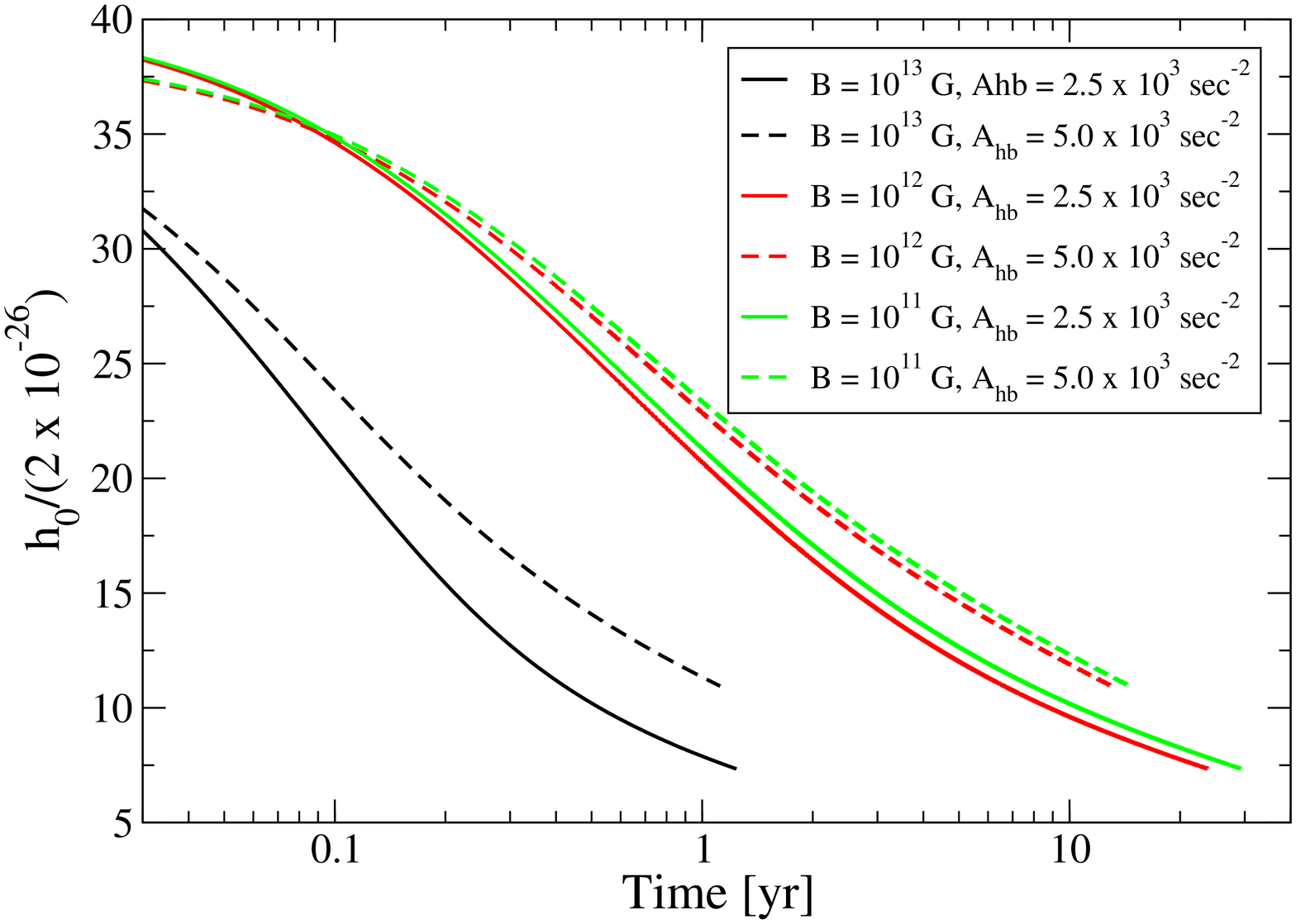}
\epsfxsize=250pt
\epsfbox{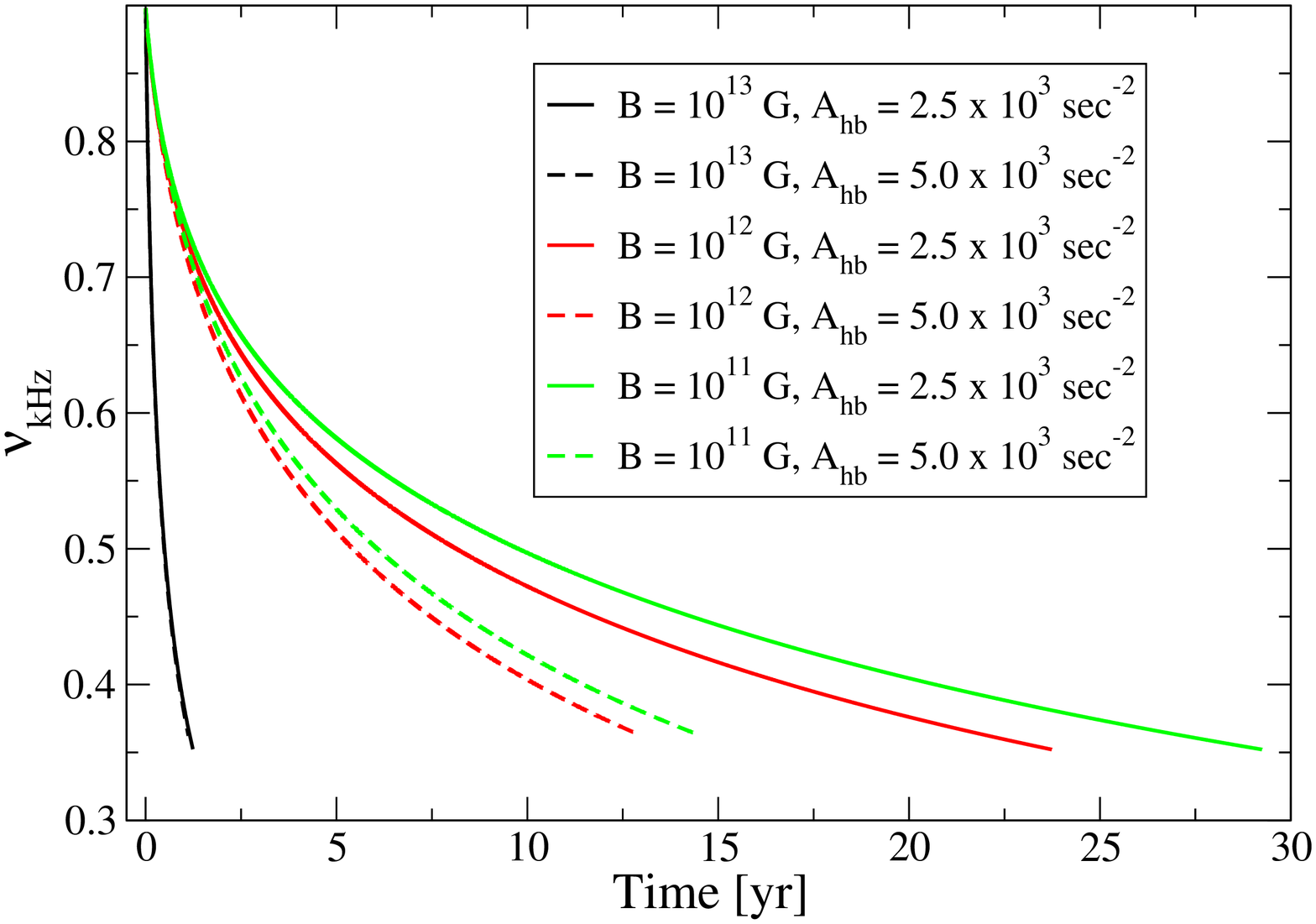}
\caption{{\bf Type I evolution.} $T_{\rm h} = 2 \times 10^9$ K. Gravitational wave amplitude at $d = 100$ kpc and spin frequency versus time for different values of the magnetic field and of $A_{\rm hb}$. }
 \renewcommand{\arraystretch}{0.75}
 \renewcommand{\topfraction}{0.6}
\label{h0fig}
\end{center}
\end{figure}

We write the gravitational wave amplitude $h_0$ following Watts {\it et al.} \cite{Watts} as
\begin{equation}
h_0^2 = \frac{5 G}{2 \pi^2 c^3 d^2 \nu_{GW}^2} \dot{E}_{GW},
\label{GWamp}
\end{equation}
where $d$ is the distance to the source and the GW frequency  $\nu_{GW} = 4 \nu/3 = (4/3)\times \Omega/(2 \pi)$. Here $\nu$ and $\Omega$ are the spin frequency and angular velocity of the star. 

 The gravitational wave amplitude can be written in terms of the r-mode amplitude using $\dot{E}_{GW} = - (\omega_\alpha/m) \dot{J}_{GW} = 2 \gamma_{GR} M R^2 |C_\alpha|^2 \Omega$ as

\begin{equation}
h_0^2 = \frac{5 G \gamma_{GR} M R^2 |C_\alpha|^2 \Omega}{\pi c^3 d^2 \nu_{GW}^2}.
\end{equation}

For our model
\begin{equation}
h_0 \approx 1.0 \times 10^{-22} \nu_{\rm kHz}^{5/2} \left(\frac{100 \, \rm kpc}{d}\right) M_{1.4} R_{12.5}^3 |C_\alpha|.
\end{equation}

The r-mode amplitude can be approximated by its quasi-stationary solution
\begin{eqnarray}
&& h_0 \approx 1.0 \times 10^{-22} \nu_{\rm kHz}^{5/2} M_{1.4} R_{12.5}^3   \left(\frac{100 \, \rm kpc}{d}\right) \\ \nonumber
&& \times \left(\frac{|C_\alpha|}{|C_\alpha|_{\rm quasi}} \right) \sqrt{\frac{\tilde \gamma_\beta \tilde \gamma_\gamma \Omega_c} {4 \tilde{\kappa}^2 \tilde \omega_\beta \tilde \omega_\gamma \tilde{\Omega}}} \\ \nonumber 
&& \times \sqrt{ 1+  \left(\frac{\tilde \Omega |\delta \tilde \tilde \omega|}{\tilde \gamma_{\alpha\, v}+\tilde \gamma_\beta +\tilde \gamma_\gamma - \tilde \gamma_{GR}}\right)^2}.
\end{eqnarray}

In type I evolutions, we can use the strong viscosity limit for the quasi-steady mode amplitude
\begin{eqnarray}
&& h_0 \approx 1.0 \times 10^{-22} \nu_{\rm kHz}^{5/2} M_{1.4} R_{12.5}^3   \left(\frac{100 \, \rm kpc}{d}\right) \\ \nonumber
&& \times \left(\frac{|C_\alpha|}{|C_\alpha|_{\rm quasi}} \right) \sqrt{\frac{\tilde \gamma_\beta \tilde \gamma_\gamma \Omega_c} {4 \tilde{\kappa}^2 \tilde \omega_\beta \tilde \omega_\gamma \tilde{\Omega}}},
\end{eqnarray}
and
\begin{eqnarray}
&& h_0 \approx 3.0 \times 10^{-24} \nu_{\rm kHz}^{2} M_{1.4} R_{12.5}^3   \left(\frac{100 \, \rm kpc}{d}\right) \\ \nonumber
&& \times \left(\frac{|C_\alpha|}{|C_\alpha|_{\rm quasi}} \right) \left(\frac{1.25}{\tilde{\kappa}}\right) \sqrt{\frac{0.38 \times 0.28}{\tilde \omega_\beta \tilde \omega_\gamma}}  \sqrt{\frac{\gamma_\beta }{1/\rm{sec}} \frac{\gamma_\gamma}{1/\rm{sec}}}.
\end{eqnarray}
 $\gamma_\beta \sim \gamma_\gamma \sim 1/$sec is typical at $\nu \sim 1$ kHz. Note that at   $\nu \sim 1$ kHz $\gamma_{GR} \approx 0.05$/sec and $\delta \omega \sim 1/$sec. The gravitational wave amplitude is independent of detuning and is directly proportional to the internal viscous dissipation. So, in principle, if fast spinning young neutron stars exist and have many modes with high viscosity that are near-resonant with the r-mode, their gravitational wave signal could provide a measurement of internal neutron star physics. 
   
   The type I evolution considered in Sec.\ \ref{type1}  ($T_h = 2 \times 10^9$ K) is the most optimistic scenario. It has a gravitational wave amplitude of $h_0 \approx 5 \times 10^{-25}$ at $\nu = 400$ Hz. All values of $h_0$ reported in this section assume a distance to the source of $d = 100$ kpc unless explicitly stated otherwise. The viscous damping rates  vary between $0.22$ sec$^{-1}$ at 900 Hz and $1.26$~sec$^{-1}$ at 300 Hz for $\gamma_\beta$, and  between $0.17$ sec$^{-1}$ and $0.92$ sec$^{-1}$ for $\gamma_\gamma$. The strong viscosity limit is accurate within 10\% of the full quasi-stationary solution for $\nu \le 500$ Hz. This evolution is independent of $\delta \omega$. 
   
   Fig.\ \ref{h0fig} shows the gravitational wave amplitude and the spin frequency as a function of time for the $T_h = 2 \times 10^9$ K type I evolution. Lowering the magnetic field can increase the duration of the spin-down from $1.1$ yr up to about $13$ yr at $B \approx 10^{12}$ G and $14.5$ yr at  $B \approx 10^{11}$ G. The spin-down timescale is approximately $\propto A_{hb}^{-1}$ in the limit when gravitational radiation spin-down torque is larger than the magnetic dipole spin-down. A more detailed discussion on how the duration of the instability scales with viscosity is presented in Appendix E.
        
  The mixed I-II evolution with $T_h = 6 \times 10^9$ K (Sec.\ \ref{type1}) had higher damping rates: $\gamma_\beta$ between $2.2$ sec$^{-1}$ (at 900 Hz) and $6.7$ sec$^{-1}$ (at 600 Hz) and $\gamma_\gamma$ is between $1.7$ sec$^{-1}$ and $5.1$ sec$^{-1}$ leading to $h_0\approx 6 \times 10^{-24}$ at $\nu = 600$ Hz. The strong damping limit is accurate within 5\% for this scenario. The type I part of the evolution lasts only about a month and in this period the spin frequency decreases from $900$ Hz to $600$ Hz. The timescale is independent of  the detuning.   Although $h_0$ is large, the rapid decrease in spin frequency over a relatively short time would make detection challenging.
  
 At about $\nu \sim 600$ Hz the star intersects the stability curve and cools across the stable region at constant spin frequency. It then reaches equilibrium on the low $T$ branch of the stability curve. This corresponds to a type II behavior and lasts longer ($\sim 1.2$ yr at $B = 10^{13}$ G), but leads to a much lower gravitational wave amplitude.  
 
 In type \cal{II} evolutions the daughter modes are not significant and the spin-down occurs in thermal
equilibrium $\dot{E}_{GW} = L_\nu(T)$, where $T=T_{\rm CFS}(\Omega)$ is determined by the CFS  stability curve from equating $\gamma_{GR} = \gamma_{\alpha\, v}$. The gravitational wave amplitude becomes
\begin{equation}
h_0^2 = \frac{5 G L_\nu(T)}{2 \pi^2 c^3 d^2 \nu_{GW}^2},
\end{equation}
Assuming direct URCA dominates the cooling 
\begin{equation}
h_0^2 \approx \frac{5 G L_{dU} T_9^6 f_{\rm dU} R_{\rm dU}(T/T_c)}{2 \pi^2 c^3 d^2 \nu_{GW}^2}.
\end{equation}
Taking the square root
\begin{eqnarray}
h_0 &\approx& \frac{T_9^3}{\nu_{GW} \, d} \sqrt{\frac{5 G L_{dU} f_{\rm dU} R_{\rm dU}(T/T_c)}{2 \pi^2 c^3}}\\ \nonumber
 &\approx& \frac{3 T_9^3}{4 \nu \, d} \sqrt{\frac{5 G L_{dU} f_{\rm dU} R_{\rm dU}(T/T_c)}{2 \pi^2 c^3}}
\end{eqnarray}
Plugging in some fiducial values gives
\begin{eqnarray}
\label{type2h}
h_0 &\approx& 1.9 \times 10^{-26} \frac{T_9^3 R_{\rm dU}}{0.01} \frac{1}{\nu_{\rm kHz}}\left(\frac{100\, \rm{kpc}}{d}\right)  \\ \nonumber
&& \times \left(\frac{L_{\rm dU}}{10^{46} \rm{erg \, sec^{-1}}}  \frac{f_{\rm dU}}{0.10} \right)^{1/2} 
\end{eqnarray}
The gravitational wave amplitude $h_0$ in this type of evolution is independent of detuning and would be a direct measurement of the strength of the neutrino cooling.

In mixed I-II evolutions the type II part of the evolution lasts the longest, $\sim 1.2$ yr for $T_h = 6 \times 10^9$ K evolution with $B = 10^{13}$ G presented in Sec.\ \ref{type1}. In this case Eq.~(\ref{type2h}) gives a gravitational wave amplitude decreasing from $h_0 \approx 5 \times 10^{-26}$ when the star first intersects the r-mode stability curve to $ h_0 \approx 2 \times 10^{-27}$ at the end of the evolution. The full type II evolution with $T_h = 6 \times 10^9$ K occurs on approximately the same spin-down timescale $\sim 1.2$ yr and has a similar $h_0$ ranging between $\approx 6 \times 10^{-26}$ and $2 \times 10^{-27}$ for the same distance of 100 kpc. 

 In mixed II-III evolutions, the type II part of the evolution is very short. It lasts only about one week for the $T_h = 1.2 \times 10^{10}$ K example in Sec.\ \ref{type2} and leads to larger $h_0 \approx 3-5 \times 10^{-24}$. In this short time the spin of the star changes by more than 30~\% of the initial spin frequency, which would make the detection difficult. This spin-down timescale for this part of the evolution is independent of detuning and also independent of the strength of the magnetic field for $B \lesssim 10^{13}$ G.
 
 In the limit when
\begin{equation}
\left(\frac{\tilde \Omega |\delta \tilde \omega|}{\tilde \gamma_{\alpha\, v}+\tilde \gamma_\beta +\tilde \gamma_\gamma -\tilde \gamma_{GR}}\right)^2  >> 1
\label{testxx}
\end{equation}

\begin{eqnarray}
\noindent && h_0 \approx 4.8 \times 10^{-25} \nu_{\rm kHz}^3 M_{1.4} R_{12.5}^3   \left(\frac{100 \, \rm kpc}{d}\right) \\ \nonumber
\noindent &&\times \left(\frac{|C_\alpha|}{|C_\alpha|_{\rm PIT}}\right) \frac{\sqrt{\gamma_\beta \gamma_\gamma}} {\gamma_{\alpha\, v}+\gamma_\beta +\gamma_\gamma - \gamma_{GR}} \left(\frac{|\delta \tilde \omega|}{2 \times 10^{-4}}\right)  \\ \nonumber
&& \times \left(\frac{1.25}{\kappa}\right)  \sqrt{\frac{0.38 \times 0.28}{\tilde{\omega}_\beta \tilde{\omega}_\gamma}}.
\end{eqnarray}
The ratio $\sqrt{\gamma_\beta \gamma_\gamma}/ (\gamma_{\alpha\, v}+\gamma_\beta +\gamma_\gamma - \gamma_{GR})$ is of order 1. In the type III evolution presented in Sec.\ \ref{type3} ($T_h = 1.2 \times 10^{10}$ K ) this ratio varies between $0.5$ and $1.3$. The spin-down time changes with detuning, magnetic field and slippage factor. For $B=10^{13}$~G and $S_{\rm ns} \approx 0.3$, $t_{\rm spin-down} \sim 1$ yr with $h_0 \sim 2 \times 10^{-26}$ at $\nu = 300$~Hz. Lowering the magnetic field to $B=10^{12}$~G leads to a spin-down time of $\sim 74$ yr for the same slippage factor. Lowering the slippage factor lowers spin frequency at which the star intersects the r-mode stability and can increase the instability timescale significantly. For $S_{\rm ns} \approx 0.005$ ($\nu_{f} \sim 80$ Hz) and $B=10^{13}$ G, the spin-down time is  $\sim 27$ yr. At $B=10^{12}$~G,
$t_{\rm spin-down} \approx 1, 400$ yr. This timescale also changes with detuning. For $|\delta \tilde \omega| \approx 2 \times 10^{-5}$ the spin-down occurs on a different $C=H$ curve at a lower temperature. For $S_{\rm ns} \approx 0.3$ the spin-down time is $\sim 75$ yr at $B=10^{12}$ G.
As $\delta \omega \to 0$ this scenario disappears. 

 An optimistic minimum detectable gravitational wave amplitude $h_0$  is given by Watts {\it et al.} \cite{Watts} \begin{equation}
h_0 = 11.4 \sqrt{\frac{S_n}{D T_{\rm obs}}},
\end{equation}
where $T_{\rm{obs}}$ is the observation time,  $D$ is the number of detectors and $S_n$ is the power spectral density of the detector noise. An integration time of about two weeks leads leads to a gravitational wave amplitude of 
\begin{eqnarray}
h_0 &=& 2 \times 10^{-26} \left(\frac{S_n}{10^{-47}\, \rm sec}\right)^{1/2} \left(\frac{2}{D}\right)^{1/2} \\ \nonumber
&& \times \left(\frac{1.21 \times 10^6\, \rm sec}{T_{\rm obs}}\right)^{1/2}
\end{eqnarray}
for Advanced LIGO. The power spectral density $S_n$ of Ref. \cite{Watts} varies between $1.4 \times 10^{-47}$ at 300 Hz and $2.2 \times 10^{-47}$ at 900 Hz for their Advanced LIGO broad band curve.

 The most optimistic scenarios for gravitational wave detection are type I evolutions with $h_0 = 5 \times 10^{-26} (1 \rm Mpc/d) (\nu/400 Hz)^2$.  The gravitational wave amplitude decreases with internal viscous dissipation. For a magnetic field of $B = 10^{13}$ G the r-mode instability timescale is $\sim 1$ yr. The timescale increases to $\sim 14$ yr when the magnetic field is lowered to $B = 10^{11}$ G and can be extended further by lowering the dissipation (See Fig.\ \ref{h0}.) In general, the timescale on which the r-mode instability is active is a combination of the gravitational spin-down and the magnetic dipole spin-down timescale.  The gravitational spin-down timescale  due to the r-mode is inversely proportional to the viscous dissipation, while the magnetic dipole spin-down timescale is approximately independent of viscosity and $\propto B^{-2}$. The An approximate analytic formula for the spin-down timescale is derived in Appendix E.  

The core-collapse supernova rate in our galaxy is still uncertain. Diehl {\it et al.} \cite{Diehl2006} estimate a galactic core collapse supernova rate of $1.9 \pm 1.1$/century from the measurements of $\gamma$-ray radiation from $^{26}$Al. The supernova formation rate within a galaxy is typically taken to be proportional to the blue band luminosity of the galaxy \cite{Capperllaro99}.  Since our galaxy has a blue band luminosity about 5 times larger than that of any other galaxy within 100 kpc, the supernova rate within this distance is approximately the same as the galactic rate (See the data from Table I of Kopparapu {\it et al.} \cite{ravi08} for blue band luminosity of galaxies up to 10 Mpc.) Advanced LIGO may see type I evolutions up to 1 Mpc. This distance would include about three times as many supernova as in our galaxy because 1 Mpc includes the Andromeda galaxy, which has more than twice the blue band luminosity of the Milky Way \cite{ravi08}. 

If a core collapse supernova occurs within 100 kpc while Advanced LIGO is on, then the gravitational radiation amplitude associated with the r-mode instability ought to be large enough to detect in all scenarios we have considered. The main complication for data analysis would be the relatively rapid, and possibly irregular, neutron star spin-down, so that the source is not strictly periodic.  In the most optimistic scenario we find, we estimate that Advanced LIGO would detect gravitational waves from r-modes up to a distance of up to 1 Mpc if the neutron star was formed less than 15 years ago. 
No supernova has been observed within 100 kpc or even 1 Mpc during the past 15 years. Thus, unless a core collapse supernova has occurred in a region that is highly obscured from Earth, we cannot draw any conclusions from the absence of LIGO sources, even in our most optimistic scenarios for gravitational radiation associated with the r-mode instability. Ultimately, if LIGO operates for decades, the probability of a core collapse supernova within the galaxy is significant.
Assuming that our galaxy has a rate of 1 SN/century, we would see 1 SN/century within 100 kpc and 3 SN/century within 1 Mpc.  If SN 1987A had produced a fast spinning pulsar, Advanced LIGO would only be able to see gravitational waves emitted by its $n=3$, $m=2$ r-mode if the instability timescale is long enough, at least 27 years to be detected by Advanced LIGO.  For our most optimistic scenario, a spin-down timescale of $\sim 30$ yr occurs for $B = 10^{11}$ G or lower when we decrease the hyperon bulk viscosity by a factor of two. However, it is unlikely that SN 1987A produced a fast spinning neutron star, as electromagnetic energy emitted by the pulsar would alter the supernova light curve at late times. Observations to date put upper limits on the luminosity from a central object in SN 1987A \cite{Xray, Hubble}. 

 \section{Limitations of the Model}
\label{limitations}
There are many ways to improve our calculation. We use the eigenmodes and eigenfrequencies of an incompressible perfect fluid star. These eigenmodes have the advantage that they are known analytically and can be expressed in terms of Legendre functions labelled by integers $n$ and $m$. This simplifies the computation of bulk$^3$ and boundary layer viscosity and allows us to use the coupling coefficients computed by Brink {\it et al.}\cite{JeandrewThesis}. As stated earlier, the inertial modes that comprise the lowest parametric instability threshold change with angular velocity. We are using an effective three mode system with typical coupling coefficient and detuning. A more realistic treatment of the modes would include higher order rotational effects in the mode frequencies and would model the change in frequencies and coupling coefficients with angular velocity. This would allow for mode changes in the lowest parametric instability threshold and the spin-down of the star could be followed more accurately. Additionally, one could include differential rotation \cite{Rezolla} and mixtures of superfluids or of superfluid and normal fluid regions \cite{superfluid}. Dissipation rates, particularly from bulk viscosity, depend on the composition of high density nuclear matter which could differ from what we assume.

\footnotetext[3]{Note that incompressible stars do not have bulk viscosity. We use an approximation to 
leading order in the adiabatic index that agrees within a factor of two with the $n=1$ polytrope calculation. See Appendix B for more details.} 

We do not include effects due to buoyant forces or magnetic fields in our mode treatment. The magnetic field frequency corrections are expected to be small. Morsink and Rezania \cite{SharonRezania} computed this frequency shift $\Delta \omega/\omega$ perturbatively for $n=m+1$ r-modes by assuming that the modes remain unchanged. Following their method we compute the shift for several high $n$ inertial modes for a constant magnetic field (See Appendix C.) We find that for a magnetic field of $B = 10^{12}$ G and a spin frequency of 1 kHz the inertial mode frequency shift is $\Delta \omega/\omega \sim B_{12}^2 (10^{-9} -10^{-8})/\nu_{kHz}^2$. This change in frequency is about three orders of magnitude larger than that for the $n=3$, $m=2$ r-mode, but still much smaller than the typical detuning of $\delta \omega/2 \Omega = 10^{-4}$. So, magnetic field corrections can be ignored. 
 
Buoyant forces are expected to change the inertial mode spectrum at low frequencies.  Roughly speaking, the Coriolis effect dominates buoyancy when the angular velocity of the star $\Omega$ is much larger than the Brunt-$\rm{V\ddot{a}is\ddot{a}l\ddot{a}}$ angular frequency $N$. When $2 \Omega > N$ the frequency spacing in the inertial mode sector (inertial mode frequencies $\omega$ lie between $-2 \Omega < \omega < 2 \Omega$ ) will not be affected significantly \cite{AP}. Passamonti {\it et al.\ }\cite{AP} find that the Coriolis force is the dominant restoring force for all modes when $\Omega > 0.3 \sqrt{\pi G \bar{\rho}}$.  The $n=m+1$ r-modes are unaffected by buoyancy at all frequencies \cite{AP}. As the frequency decreases below this value the inertial modes behave more like g-modes. As $\Omega \to 0$,  the inertial mode frequency is no longer $\propto \Omega$ and tends to a constant, leading to the loss of resonances with the r-mode.   Consequently, for a typical Brunt-$\rm{V\ddot{a}is\ddot{a}l\ddot{a}}$ spin frequency of $150$ Hz \cite{Lai1999,RG1992} the r-mode will lose its near-resonant modes roughly when $2 \Omega \le N$ ($\nu \le 75$ Hz), and in the absence of viscosity it would grow unabated. However, in most models the viscous damping rate dominates the gravitational driving at such low frequencies. For the simple viscosity model we consider in this paper the minimum spin frequency on the stability curve is higher than 75 Hz even for low slippage factors $S_{\rm ns} \sim 0.01$ and a relatively high critical temperature for hyperons of $T_h \sim 10^{10}$ K. See Sec.\ \ref{buoyancy} for a more detailed discussion. Thus, buoyancy can be safely neglected in our calculations.
%Once the neutrino cooling is stopped by viscous heating, the star spins down oscillating between thermal equilibrium states. Depending on where this balance happens in the $\Omega-T$ plane, it can spin down in different regions delimited by the r-mode stability curve.
%Some variation of the behavior we find here should be generic.
 % The evolutions
 %should still be determine by the competition between neutrino cooling and viscous heating. Depending on when the cooling can be stopped by viscous heating, one ought to find that the star spins-down in different regions delimited by the r-mode stability curve and oscillates between thermal equilibrium states. 
\section{Conclusion}
\label{conclusion}
This paper is the first treatment of the r-mode instability that uses a physical model for nonlinear saturation in newborn neutron stars. We use one triplet of modes: the $n=3, m=2$ r-mode, and two near-resonant inertial modes that couple to it. Nonlinear effects become important when the r-mode amplitude grows above its parametric instability threshold. This threshold 
provides a physical cutoff to the r-mode instability by energy transfer to other inertial modes in the system.

The behavior we find is richer than expected from previous work. We find a variety of evolutions that depend on internal neutron star physics, which is still very uncertain. The competition between viscous heating and neutrino cooling plays the most prominent role. Once the cooling is stopped by dissipative heating,  the star spins down oscillating around quasi-steady thermal equilibrium. 

Our model has some schematic aspects. We use typical detuning and coupling coefficients that will change as 
the model becomes more sophisticated. The detuning is roughly inversely proportional to the number of direct couplings to the r-mode, which is a function of the number of modes included. Some evolutions are independent of this frequency difference between modes. When the inertial mode viscosity is larger than the detuning, the parametric instability threshold becomes independent of the detuning.  The threshold in the strong viscosity limit coincides with its value for zero detuning among the frequencies of the mode triplet and still leads to significant nonlinear effects.

Some of our evolutions lead to a gravitational radiation amplitude that is detectable by Advanced LIGO. 
Assuming a broad band Advanced LIGO curve with two detectors and an integration time of about two weeks, we find that gravitational radiation would be detectable within about 1 Mpc of Earth. The sources would have to be young neutron stars within years or perhaps decades after formation. The gravitational wave amplitude and the duration of the emission
depend on the internal dissipation of the modes of the star. Significant spin-down of the star over short periods of time ($\sim$ weeks) would make detection challenging. However, if fast spinning young neutron stars exist and are detected, the gravitational wave amplitude could give unique information on internal neutron star physics.

\section*{Acknowledgments}
We thank Sharon Morsink and Jeandrew Brink for useful discussions in understanding the effect of high spin frequencies on coupling coefficients and detuning. We are grateful to Badri Krishnan for providing us with the Advanced LIGO and Einstein Telescope noise curves from their paper  \cite{Watts}. We thank
David Shoemaker for another Advanced LIGO noise curve.
%and to David Shoemaker for another LIGO noise not used here. 
R.B. is grateful to Jayashree Balakrishna and Gregory Daues for hospitality during her stay in St. Louis for the April APS meeting, discussions, and their constant friendship and support.  She also thanks
Andrew Lundgren for support and useful discussions.

This research was funded by NSF grants AST-0606710 and PHY-0652952 at Cornell University. RB also acknowledges the support of National Science Foundation Grant No.~PHY 06-53462 and No.~PHY 05-55615, and NASA Grant No.~NNG05GF71G, awarded to The Pennsylvania State University.

\section*{Appendix A: Quasi-Steady Mode Amplitudes} 

In terms of amplitudes and phase variables $C_j = |C_j| e^{i \phi_j}$ Eqs.\ (\ref{eqcode}) can be rewritten as
\begin{eqnarray}
\label{phase}
\frac{d |\bar{C}_\alpha|}{d\tilde{\tau}} &=&\frac{ \tilde{\gamma}_\alpha}{\tilde{\Omega} |\delta \tilde{\omega}|} |\bar{C}_\alpha| - \frac{\sin \phi |\bar{C}_\beta| |\bar{C}_\gamma|}{2 \sqrt{\tilde{\Omega}}} ,  \\  \nonumber
\frac{d |\bar{C}_\beta|}{d\tilde{\tau}} &=& - \frac{ \tilde{\gamma}_\beta}{\tilde{\Omega} |\delta \tilde{\omega}|} |\bar{C}_\beta| + \frac{\sin \phi |\bar{C}_\alpha| |\bar{C}_\gamma|}{2 \sqrt{\tilde{\Omega}}} , \\  \nonumber
\frac{d |\bar{C}_\gamma|}{d\tilde{\tau}} &=& - \frac{ \tilde{\gamma}_\gamma}{\tilde{\Omega} |\delta \tilde{\omega}|} |\bar{C}_\gamma| +\frac{\sin \phi |\bar{C}_\alpha| |\bar{C}_\beta|}{2 \sqrt{\tilde{\Omega}}} , \\  \nonumber
\frac{d \phi}{d \tilde{\tau}} &=&  \frac{\delta \tilde{\omega}}{|\delta \tilde{\omega}|}  - \frac{\cos\phi}{2 \sqrt{\tilde \Omega}} \left(\frac{|\bar{C}_\beta| |\bar{C}_\gamma|}{|\bar{C}_\alpha|}-\frac{|\bar{C}_\alpha| |\bar{C}_\gamma|}{|\bar{C}_\beta|} - \frac{|\bar{C}_\beta| |\bar{C}_\alpha|}{|\bar{C}_\gamma|} \right) ,
\end{eqnarray}
where we have defined the relative phase difference as $\phi = \phi_\alpha - \phi_\beta - \phi_\gamma$.

The quasi-stationary solution in Eq.\ (\ref{stationarySol}) is found by setting $d |C_j|/d \tilde \tau = 0$ and $d\phi/d\tilde \tau =0$ simultaneously.
%\begin{eqnarray}
%\label{stationarySolApp}
%|\bar{C}_\alpha|^2 = \frac{4 \tilde{\gamma}_\beta \tilde{\gamma}_\gamma}{\tilde{\Omega} |\delta \tilde{\omega}|^2} \left(1 + \frac{1}{\tan^2 \phi}\right) , \\ \nonumber
%|\bar{C}_\beta|^2 = \frac{4 \tilde{\gamma}_\alpha \tilde{\gamma}_\gamma}{\tilde{\Omega} |\delta \tilde{\omega}|^2} \left(1 + \frac{1}{\tan^2 \phi}\right) , \\ \nonumber
%|\bar{C}_\gamma|^2 = \frac{4 \tilde{\gamma}_\alpha \tilde{\gamma}_\beta}{\tilde{\Omega} |\delta \tilde{\omega}|^2} \left(1 + \frac{1}{\tan^2 \phi}\right) , \\ \nonumber
%\tan \phi = \frac{\tilde{\gamma}_\beta + \tilde{\gamma}_\gamma - \tilde{\gamma}_\alpha}{\tilde{\Omega} |\delta \tilde{\omega}|} .
%\end{eqnarray}

\section*{Appendix B: Viscous Dissipation}
\label{AppBlayInt}
This appendix discusses the dissipation we use for the r-mode that has mode index
$j=4$ ($n=3$, $m=2$) and two near resonant inertial modes with
$j = 494$ ($n=14$, $m=-5$) and $j=592$ ($n=15$, $m=3$).
%%$j = 414$ ($n=13$, $m=-3$) and $j=538$ ($n=14$, $m=1$) for the lowest threshold
Following Brink {\it et al.} \cite{Jeandrew2}, we label the hybrid inertial modes so that each mode is given
a unique number $j$ that is a function of principal Legendre index $n$ ($n: 2 \to \infty$), azimuthal number 
$m$ ($m: 0 \to n-1$) and frequency index $k$ ($k: 1 \to n-m$ if $m \ne 0$ and $1 \to n-m-1$ if $m=0$).
\begin{equation}
j = \frac{(n-1)n(n+1)}{6} + \frac{(n-m-1) (n-m)}{2} + k - 1
\end{equation}
We compute the bulk viscosity using the modes of the incompressible star by taking the dissipation energy to leading order in the adiabatic index $\Gamma_1$
\begin{equation}
\dot{E}_B = - \int d^3x \frac{\zeta \omega_j^2}{\Gamma_1^2} \left| \frac{{\bf \xi \cdot \nabla} p}{p}\right|^2,
\end{equation}
and setting $\Gamma_1 = 2$. This approximation was proposed by Cutler and Lindblom \cite{CL} and adopted by
Kokkotas and Stergiouluas \cite{KS} for the r-mode and by Brink {\it et al.} \cite{Jeandrew1, Jeandrew2, Jeandrew3, JeandrewThesis} for inertial modes.  

Table \ref{tabBulk} compares the bulk viscosity timescales for several different inertial modes
of $n=1$ polytrope computed by Lockitch and Friedman \cite{KL} with those computed 
 for an incompressible model. The difference is small;  typically about a factor of two or less.
For the computations in this table we used n-p-e bulk viscosity with a bulk viscosity coefficient 
\begin{equation}
\zeta = 6 \times 10^{25} \left(\frac{\rm{Hz}}{\omega}\right)^2 \left(\frac{\rho}{10^{15} \; \rm{g \, cm^{-3}}}\right)^2 T_9^6 \; \rm{g\, cm \, sec^{-1}}.
\end{equation}
\begin{table}[h]
\caption{Bulk viscosity timescales computed using incompressible stellar modes  ($\tau_{{\rm bulk} \, n=0}$; used in this work with a different bulk viscosity coefficient) and $n=1$ polytrope ($\tau_{{\rm bulk} \, n=1}$) inertial modes for an n-p-e gas. The $n=1$ polytrope calculation was performed by Lockitch and Friedman \cite{KL}. Note that, mathematically, the bulk viscosity damping rate is zero for incompressible stars. We adopt the attitude that the dissipation timescale is computed to leading order in $\Gamma_1$ and take $\Gamma_1 =2$ (for incompressible stars $\Gamma_1 \to \infty$). The timescales for the two models are roughly within a factor of 2 of each other.}
\begin{tabular}{|c|c|c|c|c|}
\hline
j &$n$&$m$&$\tau_{{\rm bulk}\, n=0}$& $\tau_{{\rm bulk}\, n=1}$ \\ \hline
12&4&2& $7.03 \times 10^9$ sec & $3.32 \times 10^9$ sec \\
14&4&1 &$9.68 \times 10^9$ sec & $5.86 \times 10^9$ sec \\
44&6&2&$7.00\times 10^9$ sec & $4.79 \times 10^9$ sec \\
47&6&1&$2.51 \times 10^9$ sec&$2.57 \times 10^9$ sec \\
70&7&2&$6.62 \times 10^9$ sec & $5.32 \times 10^9$ sec \\
\hline
\end{tabular}
\label{tabBulk}
\end{table}

For hyperon bulk viscosity we take the viscous damping rates to be
\begin{eqnarray}
\gamma_{\alpha \; \rm{hb}} &=&  \frac{\tau(T,T_h) A_{\rm{hb}} \tilde{\Omega}^4 I_\alpha}{1+(\tilde{\omega}_\alpha \Omega \tau(T,T_h))^2} \, \rm{sec}^{-1}, \\ \nonumber
\gamma_{j \; \rm{hb}} &=&  \frac{\tau(T,T_h) A_{\rm{hb}} \tilde \omega_j^2 I_j}{1+(\tilde{\omega}_j \Omega \tau(T,T_h))^2} \, \rm{sec}^{-1}.
\end{eqnarray}
The mode integrals $I_{{\rm hb} \, j}$ for the inertial modes is computed using
\begin{equation}
\label{Ihb}
I_{{\rm hb} \, j} =\frac{1}{\Gamma_1^2 R^3} \int d^3x \left|\frac{{\bf\xi}_j \cdot \nabla p}{p} \right|^2, 
\end{equation}
The r-mode integral is computed by fitting results of Nayyar and Owen \cite{mohit}.

We treat the critical temperature $T_h$ and the coefficient $A_{\rm{hb}}$ as parameters.  $A_{\rm{hb}} = 5 \times 10^3$ sec$^{-2}$ is chosen so that the peak of the r-mode stability curve is at roughly 1000 Hz for all our stable evolutions. We lower it to exhibit runaway behavior.

 For the two inertial modes of interest 
\begin{eqnarray}
%I_{414} &\approx& 2 \pi \times 50.3, \;\;\;\;\; I_{538} \approx 2 \pi \times 100.1, \\ \nonumber
I_{{\rm hb} \, 592} \approx \frac{2 \pi \times 103.1}{\Gamma_1^2}, \;\;\;\;\; I_{{\rm hb} \, 494} &\approx& \frac{2\pi \times 142.4}{\Gamma_1^2}.
\end{eqnarray}
For the r-mode
\begin{equation}
I_{{\rm hb} \, 4} = I_{{\rm hb} \alpha} = 0.211.
\end{equation}
The relaxation timescale

\begin{equation}
\tau(T,T_h) = \frac{t_1 T_9^{-2}}{R_{\rm{hb}}(T/T_h)},
\end{equation}
where we use the reduction factor proposed by Ref. \cite{HLY} 
\begin{equation}
R_{\rm{hb}}(T/T_c) = \frac{a^{5/4} + b^{1/2}}{2} \exp\left(0.5068 - \sqrt{0.5068^2+y^2}\right)
\end{equation}
where $a = 1+ 0.3118 y^2$,  $b = 1 + 0.2566 y^2$ and $y = \sqrt{1.0 - T/T_h} (1.456 - 0.157 \sqrt{T_h/T} + 1.764 T_h/T)$.
The constant is taken to be $t_1 \approx 10^{-6}  \; \rm{sec}$.

The boundary layer viscosity is computed via Eq.\  (3) of Bildsten and Ushomirsky \cite{BU}
and is given by
\begin{eqnarray}
\gamma_{j \; \rm{bl}}(T,\Omega) =  I_j A_{\rm bl}  \tilde{\omega}^{5/2} S_{\rm{ns}}^2 \frac{\sqrt{\tilde{\Omega}}}{T_9}
\end{eqnarray}
 where $S_{\rm ns}$ is the slippage factor and $I_j$ is the mode integral for mode $j$.  The slippage factor is the fractional difference in velocity between the crust and core of the star \cite{LU, Wagoner}. Here the constant
 \begin{equation}
 A_{\rm bl} = 3.68 \times 10^{-5}  \left(\frac{\rho}{\rho_b}\right) \sqrt{\frac{R_{12.5}}{M_{1.4}}}  \, \rm{sec}^{-1}
 \end{equation}
 and the mode integral 
 \begin{equation}
 \label{Ibl}
 I_{{\rm bl} \, j} = \int d\cos \theta d \phi \frac{{\bf \xi}_j \cdot {\bf \xi}_j^\star}{R^2}, 
 \end{equation}
 where $\xi_j$ is the displacement vector for mode $j$ and the density on the boundary layer is taken to be constant $\rho_b =1.5 \times 10^{14}$ g cm$^{-3}$.  In these expressions we are using the same
 normalization as Schenk {\it et al.}\cite{Schenk}: $\int d^3 x \, \rho \; {\bf \xi \cdot \xi} = M R^2$.

For the r-mode this gives
\begin{eqnarray}
I_{{\rm bl} \, 4}^{\rm incompressible} &=& 2 \pi \times 10.5,  \\ \nonumber
 I_{{\rm bl} \, 4}^{n=1 \, \rm polytrope} &\approx& 2 \pi \times 21.8,
\end{eqnarray}
where we have used the incompressible value in this paper.

For the daughter modes at first two parametric instability thresholds the integrals are calculated also using the
modes for an incompressible star
\begin{eqnarray}
%I_{414} &\approx& 2 \pi \times 92.99, \;\;\;\;\; I_{538} \approx 2 \pi \times 376.2, \\ \nonumber
I_{{\rm bl} \, 494} &\approx& 2\pi \times 241.8, \;\;\;\;\; I_{{\rm bl} \, 592} \approx 2 \pi \times 140.1.
\end{eqnarray}

\section*{Appendix C: Frequency Change due to Magnetic Fields}
\label{AppC}
In this appendix we follow Morsink and Rezania \cite{SharonRezania} to obtain the frequency corrections due to the presence
of a magnetic field. These corrections are added perturbatively.

They define a dimensionless magnetic coupling
\begin{equation}
\kappa_{AB} = M^{-1} <{\bf \xi}_A, \rho^{-1} \bf{F}_B>,
\end{equation}
where ${\bf F}$ is the Lorentz force created by the fluid motion. The coefficients can be thought of as the ratio of the work done by the perturbed Lorentz force to the total magnetic energy stored in the equilibrium star.

The magnetic coupling coefficients can be written as \cite{SharonRezania}
\begin{equation}
\kappa_{AB} =- \frac{1}{4 \pi {\cal M}} \left[ \kappa_{AB}^{(1)} - \kappa_{AB}^{(2)} - \frac{\omega_B^2}{c^2} \kappa_{AB}^{(3)}\right],
\end{equation}
where
\begin{eqnarray}
\kappa_{AB}^{(1)} &=& \int  d^3x \; \nabla \times ({\bf \xi}_A^\star \times {\bf B}) \cdot \nabla \times ({\bf \xi}_B \times {\bf B}) \\ \nonumber
\kappa_{AB}^{(2)} &=& \int d^3x\; {\bf \xi^\star_A \times (\nabla \times B) \cdot  \xi_B \times (\nabla \times B)} \\ \nonumber
\kappa_{AB}^{(3)} &=& \int d^3x \; {\bf  (\xi^\star_A \times B) \cdot  (\xi_B \times B)}. \\ \nonumber
\end{eqnarray}
 We neglect $\kappa_{AB}^{(2)}$ and $\kappa_{AB}^{(3)}$. We are considering a constant $B$ field and so $\kappa_{AB}^{(2)} = 0$. The errors introduced by neglecting $\kappa_{AB}^{(3)}$ are of the same order as those from neglecting general relativity \cite{SharonRezania}.  Assuming the off-diagonal entries are small \cite{SharonRezania},  the frequency corrections are given by
\begin{equation}
\omega_{\rm new} = \omega_{\rm old} \left(1 - \frac{\cal{M}}{2\cal{\epsilon}}  \kappa_{AA} \right),
\end{equation}
where ${\cal M}/\epsilon$ is the ratio of magnetic field energy to rotational kinetic energy. The rotational energy is $ \epsilon = M R^2 \Omega^2$ and the magnetic field energy is given by ${\cal M} = B^2 R^3/6$.

Assuming a constant magnetic field of the form
\begin{eqnarray}
B_x &=& B_0 \sin \alpha, \\ \nonumber
B_y &=& 0, \\ \nonumber
B_z &=& B_0 \cos \alpha,
\end{eqnarray}
one can easily compute the magnetic coupling using the modes of an incompressible star (Eq.\ (3.18)  together with the recursion relations Eq. (A.1-5) in \cite{JeandrewThesis}).) %In cylindrical coordinates the
%integral is 
%\begin{eqnarray}
%|\kappa_{AA}| &=& \sin^2 \alpha \int d^3x \left\{\sin^2 \phi \cos^2 \phi \left[ \left(\partial_\varpi |\xi^\varpi| + m |\xi^\phi| - \frac{|\xi^\varpi|}{\varpi}\right)^2 \right. \right. \\ \nonumber 
%&+&\left. \left. \left(\varpi \partial_\varpi |\xi^\phi| + \frac{m}{\varpi} |\xi^\varpi|\right)^2 \right]
%+\left(\varpi  \partial_\varpi |\xi^\phi|  \cos^2 \phi+ |\xi^\phi| \right. \right.\\ \nonumber
%&-& \left. \left. \frac{m}{\varpi} |\xi^\varpi| \sin^2 \phi \right)^2 + \left[\partial_z|\xi^z| + \sin^2 \phi \partial_\varpi |\xi^\varpi|  \right.\right. \\ \nonumber
%&+& \left. \left. \cos^2 \phi \left(-m |\xi^\phi| + \frac{|\xi^\varpi|}{\varpi}\right)\right]^2 
%+\cos^2 \phi (\partial_\varpi |\xi^z|)^2 \right. \\ \nonumber
%&+& \left. \sin^2\phi \frac{m^2}{\varpi^2} |\xi^z|^2 \right\} 
%+ \cos^2 \alpha \int d^3 x \left[\partial_z |\xi^\varpi|^2 + \varpi^2 (\partial_z |\xi^\phi|)^2 \right. \\ \nonumber
%&+&\left.  \left(\partial_\varpi |\xi^\varpi| 
%+ \frac{|\xi^\varpi|}{\varpi} - m |\xi^\phi|\right)^2\right]
%\end{eqnarray}
%\begin{eqnarray}
%|\kappa_{AA}| &=& \int d^3x \sin^2 \alpha \{ \cos^2 \phi (|\partial_\varpi \xi_\varpi|^2 + |\partial_\varpi \xi^\phi|^2 + |\partial_\varpi \xi_z|^2) \\ \nonumber
%&+& (\sin^2 \phi/\varpi^2) [ (1+m^2) (|\xi^\varpi|^2 + |\xi^\phi|^2) + m^2 |\xi^z|^2 - 4 m |\xi^\phi| |\xi^\varpi|]\} \\ \nonumber 
%&+& \cos^2 \alpha (|\partial_z \xi^z|^2 + |\partial_z \xi^\varpi|^2 + |\partial_z \xi^\phi|^2)
%\end{eqnarray}

For $n=m+1$ modes this computation can be performed analytically \cite{SharonRezania}
\begin{eqnarray}
|\kappa_{r-modes}| &=& \frac{4 \pi (m+1) (2 m +3)}{12}  \times \\ \nonumber 
&&\left(1 + \frac{m^2 +m - 3}{2} \sin^2 \alpha \right).
\end{eqnarray}
For the $n=3$, $m=2$ mode
\begin{equation}
|\kappa_{j=4}| \approx 22 + 33 \sin^2 \alpha
\end{equation}
For the two daughter modes used in this paper
\begin{eqnarray}
%|\kappa_{414}| &\approx& 10^3 (1.6 + 13.5 \sin^2 \alpha) \\ \nonumber
%|\kappa_{538}| &\approx& 10^3 (2.0 + 53.5 \sin^2 \alpha) \\ \nonumber
|\kappa_{494}| &\approx& 10^3 (1.8 + 42.0 \sin^2 \alpha) \\ \nonumber
|\kappa_{592}| &\approx& 10^3 (2.0 + 23.5 \sin^2 \alpha).
\end{eqnarray}
The frequency change is 
\begin{equation}
\frac{\Delta \omega}{\omega_{\rm old}} = \frac{\omega_{\rm new} - \omega_{\rm old}}{\omega_{\rm old}} = \frac{{\cal M}}{2 \epsilon} \kappa_j.
\end{equation}
For the fiducial values used here
\begin{equation}
\frac{{\cal M}}{2 \epsilon} \approx 10^{-12} \frac{B_{12}^2 R_{12.5}}{M_{1.4} \nu_{\rm kHz}^2},
\end{equation}
where $B_{12} = 10^{12}$ G.
So, the frequency change $\Delta \omega_j/\omega_j$ for the mode $j$ is 
\begin{eqnarray}
\frac{\Delta \omega_4}{\omega_4} &=& 10^{-11} \frac{B_{12}^2 R_{12.5}}{M_{1.4} \nu_{\rm kHz}^2} (2.1 + 3.15 \sin^2 \alpha), \\ \nonumber
%\frac{\Delta \omega_{414}}{\omega_{414}} &=& 10^{-9} \frac{B_{12}^2 R_{12.5}}{M_{1.4} \nu_{\rm kHz}^2} (1.6 + 13.5 \sin^2 \alpha) \\ \nonumber
%\frac{\Delta \omega_{538}}{\omega_{538}} &=& 10^{-9} \frac{B_{12}^2 R_{12.5}}{M_{1.4} \nu_{\rm kHz}^2}(2.0 + 53.5 \sin^2 \alpha)  \\ \nonumber
\frac{\Delta \omega_{494}}{\omega_{494}} &=& 10^{-9} \frac{B_{12}^2 R_{12.5}}{M_{1.4} \nu_{\rm kHz}^2} (1.8 + 42.0 \sin^2 \alpha) \\ \nonumber
\frac{\Delta \omega_{592}}{\omega_{592}} &=& 10^{-9} \frac{B_{12}^2 R_{12.5}}{M_{1.4} \nu_{\rm kHz}^2} (2.0 + 23.5 \sin^2 \alpha)
\end{eqnarray}
The frequency changes are much smaller than the typical detuning value of $\delta \omega/(2 \Omega) \sim 10^{-4}$  for the magnetic field values considered in this paper of $B \lesssim 10^{13}$ G.
%Calculation of the frequency shift for the R-modes and daughter modes due to the magnetic field.
%Some final formula may go in the text. 
\section*{Appendix D: Stability Near Thermal Equilibrium - One Mode Evolutions}
\label{AppD}
Consider the one-mode evolution equations
\begin{eqnarray}
\frac{d|C_\alpha|^2}{dt} &=& 2 (\gamma_{GR} - \gamma_{\alpha\, v}) |C_\alpha|^2 \\ \nonumber
 C(T) \frac{dT}{dt} &=& 2 M R^2 \Omega |C_\alpha|^2 \gamma_{\alpha\, v} - L_\nu (T).
\end{eqnarray}
We expand each variable to first order around its equilibrium value
\begin{eqnarray}
|C_\alpha|^2 &=& |C_\alpha|^2_e (1 + \zeta_\alpha) \\ \nonumber
T &=& T_e (1 + \zeta_T).
\end{eqnarray}
 In equilibrium 
\begin{eqnarray}
\label{Eq}
\gamma_{GR}(\Omega) - \gamma_{\alpha\,v} (\Omega,T_e)  &=& 0\\ \nonumber
2 E_{\alpha\,e}  \gamma_{\alpha\,v} (\Omega,T_e) &=& L_\nu(T_e).
\end{eqnarray}
This leads to the coupled equations
\begin{eqnarray}
\frac{d \zeta_\alpha}{dt} &=& 2 (\gamma_{GR} - \gamma_{\alpha\, v})_e \zeta_\alpha  - 2 T_e\left(\frac{\partial \gamma_{\alpha\, v}}{\partial T}\right)_e \zeta_T, \\ \nonumber
\frac{d \zeta_T}{dt} &=& \frac{2 E_{\alpha\,e} \gamma_{\alpha \,v}(\Omega,T_e)}{T_e C(T)} \zeta_\alpha   \noindent  \\ \nonumber
&&+ \left[\frac{2 E_{\alpha\,e}}{C(T)}  \left(\frac{\partial \gamma_{\alpha\, v}}{\partial T}\right)_e - \frac{1}{C(T)} \left(\frac{d L_\nu}{dT}\right)_e\right] \zeta_T,
\end{eqnarray}
where $E_{\alpha\,e} = M R^2 \Omega |C_\alpha|^2_e$.
Using Eqs.\ (\ref{Eq}) the coupled equations can rewritten as
\begin{eqnarray}
\frac{d \zeta_\alpha}{dt} &=&  - 2 T_e\left(\frac{\partial \gamma_{\alpha\, v}}{\partial T}\right)_e \zeta_T, \\ \nonumber
\frac{d \zeta_T}{dt} &=& \frac{L_\nu(T_e)}{T_e C(T_e)} \zeta_\alpha   \noindent 
+ \left[\frac{1}{\gamma_{\alpha\,v\,e}}  \left(\frac{\partial \gamma_{\alpha\, v}}{\partial T}\right)_e \right. \\ \nonumber
&& \left. - \frac{1}{L_\nu(T_e)} \left(\frac{\partial L_\nu}{\partial T}\right)_e\right] \frac{L_\nu(T_e)}{C(T_e)}\zeta_T.
\end{eqnarray}
We can now write the eigenvalue equation for this system
\begin{eqnarray}
\lambda^2 &-& \lambda \left(\frac{\partial \ln \gamma_{\alpha \, v}}{\partial T} - \frac{\partial \ln L_\nu}{\partial T}\right)_e \frac{L_\nu(T_e)}{C(T_e)} \\ \nonumber 
&+& \frac{2 L_\nu(T_e)}{C(T_e)} \left(\frac{\gamma_{\alpha \, v}}{\partial T}\right)_e = 0
\end{eqnarray}
with solutions
\begin{equation}
\lambda_{1,2} = \frac{\gamma_e}{2} \pm \frac{1}{2} \sqrt{\gamma_e^2 - \frac{8 L_\nu(T_e)}{C(T_e)} \left(\frac{\gamma_{\alpha \, v}}{\partial T}\right)_e}.
\end{equation}
Here 
\begin{equation}
\gamma_e = \left(\frac{\partial \ln \gamma_{\alpha \, v}}{\partial T} - \frac{\partial \ln L_\nu}{\partial T}\right)_e \frac{L_\nu(T_e)}{C(T_e)}.
\end{equation}
Points on the right side of the r-mode stability curve $T > T_{\rm peak}$ have a viscosity with negative slope $(\partial\gamma_{\alpha \, v}/\partial T)_e < 0$ and are always unstable (one eigenvalue is positive). While points on the left side  of the r-mode stability curve $T < T_{\rm peak}$ have $(\partial\gamma_{\alpha \, v}/\partial T)_e > 0$ are stable if $\gamma_e < 0$ and unstable if $\gamma_e > 0$.

In order to gain a better understanding of the thermal cycles around the stability curve we write the viscous heating $U$ as fraction of the cooling and subsequently study the evolution of $f$.
\begin{equation}
U  = 2 \gamma_{\alpha\, v} E_\alpha = f(t) L_\nu(T),
\label{Udef}
\end{equation}
where $f = 1$ corresponds to equal viscous heating and cooling and $f \ge 0$. We have neglected the viscous heating due to the daughter modes as their amplitudes are much smaller that that of the r-mode in this scenario. The thermal evolution of the system can now be written as
\begin{equation}
C(T) \frac{dT}{dt} = U  - L_\nu(T) = (f -1) L_\nu(T).
\end{equation}
To find the evolution of f we take the time derivative of Eq.\ (\ref{Udef}). We can then write
\begin{equation}
\label{dfdt}
\frac{1}{f} \frac{df}{dt} =\frac{1}{U} \frac{d U}{dt}- \frac{(f-1)}{C(T)} \frac{\partial L_\nu}{\partial T}
\end{equation}
and 
\begin{equation}
\label{dUdt}
\frac{1}{U} \frac{dU}{dt} = \frac{1}{\gamma_{\alpha \, v}} \frac{\partial \gamma_{\alpha \, v}}{\partial T} \frac{(f-1) L_\nu(T)}{C(T)} + 2 (\gamma_{GR} - \gamma_{\alpha \, v}).
\end{equation}
Plugging Eq.\ (\ref{dUdt}) in Eq.\ (\ref{dfdt}) we can write 
\begin{eqnarray}
\frac{1}{f} \frac{df}{dt}  &=& \frac{(f-1) L_\nu(T)}{C(T)} \left(\frac{1}{\gamma_{\alpha \, v}} \frac{\partial \gamma_{\alpha \, v}}{\partial T}-\frac{1}{L_\nu} \frac{\partial L_\nu}{\partial T} \right) \\ \nonumber
&+& 2 (\gamma_{GR} - \gamma_{\alpha \, v}).
\end{eqnarray}
Labeling the term in parenthesis as $\gamma$
we obtain
\begin{equation}
\label{dlnf}
\frac{1}{f} \frac{df}{dt}  = \frac{(f-1) L_\nu(T)}{C(T)} \gamma + 2 (\gamma_{GR} - \gamma_{\alpha \, v}),
\end{equation}
where
\begin{equation}
\gamma = \frac{1}{\gamma_{\alpha \, v}} \frac{\partial \gamma_{\alpha \, v}}{\partial T}-\frac{1}{L_\nu} \frac{\partial L_\nu}{\partial T}.
\end{equation}

Initially, the star is very hot $T\sim 10^{10}$ K and cools fast: $\gamma < 0$ (the slope of the r-mode stability curve is negative for $T>T_{\rm peak}$), $f<<1$ and $\gamma_{GR}> \gamma_{\alpha \, v}$. So, the right hand side of Eq.\ (\ref{dlnf}) is positive and $f$ grows exponentially. All fixed points on the $T> T_{\rm peak}$ side of the r-mode stability curve are unstable. So, there are no evolutions in which the star spins down along the right branch ($T>T_{\rm peak}$) of the r-mode stability curve.  The star can find thermal equilibrium $f \approx 1$ in a one mode evolution only close to the r-mode stability curve $\gamma_{GR} \approx \gamma_{\alpha \, v}$. The thermal oscillations around the stability curve can be understood from this linear perturbation analysis. Typically, the star overshoots the stability curve at first. Once this happens, the right hand side of Eq.\ (\ref{dlnf}) becomes positive ($f=1$ and $\gamma_{GR} > \gamma_{\alpha\, v}$). This makes $f$ increase and so the star heats. The neutrino cooling ($\propto T^6$) eventually balances the heating as the temperature increases. This balance happens in the stable region $\gamma_{GR}<  \gamma_{\alpha\, v}$. At this point ($f=1$ and $\gamma_{GR}<  \gamma_{\alpha\, v}$) the right hand side of  Eq.\ (\ref{dlnf}) is negative and $f$ starts decreasing (the star cools). The star enters the unstable region again and thermal oscillation repeats. In this time the angular velocity of the star decreases slowly. So, the next oscillation will have a lower amplitude.

In other words, the thermal equilibrium points at fixed $\Omega$ on the left side of the  r-mode stability curve ($T< T_{\rm peak}$; positive slope) are initially stable. The r-mode stability curve acts as an attractor. The star exhibits thermal oscillations at constant angular velocity around this curve with the oscillations becoming smaller and smaller until the trajectory of the star coincides with the r-mode stability curve. As the star spins down the viscosity decreases and the heating is slower. If the thermal equilibrium becomes unstable, then the thermal oscillations restart with growing amplitude until the return to thermal equilibrium is no longer possible. The star cools until the daughter modes are excited and the viscosity due to all three modes balances the cooling.
Otherwise, the star continues cooling and spinning down on the r-mode stability curve. Eventually, the star enters the stable regime
again after boundary layer viscosity dominates bulk viscosity and the slope of the r-mode stability curve changes.
\section*{Appendix E: Spin-down Timescales - Scaling with Viscosity}
For type I evolutions 
\begin{equation}
t_{\rm spin-down}^{GR} \propto \int_{\tilde\Omega_i = 0.67}^{\tilde\Omega_f} \int d\tilde\Omega \frac{\tilde\Omega^{-5}}{\tilde\gamma_\beta \tilde\gamma_\gamma}.
\label{tspin}
\end{equation}
At first, for simplicity, we set $R_{\rm hb} = R_{dU} = 1$. In this case, assuming that hyperon bulk viscosity dominates
\begin{eqnarray}
\gamma_{\alpha v} \propto A_{hb} T_9^{-2} \tilde\Omega^4,
\gamma_\gamma \propto A_{hb} T_9^{-2},
\gamma_\beta \propto A_{hb} T_9^{-2}.
\end{eqnarray}
%The $C=H$ curve for quasi-stationary amplitude solutions 
%\begin{equation} 
%\tilde\gamma_{\alpha\; v} \tilde\gamma_\beta \tilde\gamma_\gamma \propto T_9^6
%\label{CHspin}
%\end{equation}
%leads to 
%\begin{equation}
%T_9 \propto \tilde \Omega^{1/3} A_{\rm hb}^{1/4}.
%\label{scaleThb}
%\end{equation}
% Performing the integral in Eq.\ \ref{tspin}:
%\begin{equation}
%t_{\rm spin-down}^{GR} \propto (\tilde \Omega_f^{-8/3}  -\tilde \Omega_i^{-8/3}) A_{\rm hb}^{-1}.
%\label{tspin2}
%\end{equation}
%Assuming the final point is on the r-mode stability curve gives 
%\begin{equation}
%\Omega_f \propto A_{\rm hb}^{3/16}.
%\label{scaleomhb}
%\end{equation}
% The term proportional to the initial angular velocity $\tilde \Omega_i$ can be neglected and 
% so $t_{\rm spin-down}^{GR} \propto A_{\rm hb}^{-3/2}$.
 
 Assuming that $T_9$ does not change significantly, we can approximate the spin-down timescale by
 \begin{eqnarray}
 t_{\rm spin-down}^{GR} &\propto& \frac{1}{\tilde \gamma_\beta \tilde \gamma_\gamma} \int d \tilde \Omega \tilde \Omega^{-5} \\ \nonumber
 && \approx \tilde\Omega_f^{-4} T_{9f}^4 A_{\rm hb}^{-2}.
 \end{eqnarray}
 On the $C=H$ curve
 \begin{equation}
\tilde\Omega^6 A_{\rm hb}^2 T_{9}^{-4} \propto T_{9}^6.
\end{equation}
So, 
\begin{equation}
T_{9} \propto A_{\rm hb}^{1/5} \tilde\Omega_f^{3/5}
\label{scaleTAhb}
\end{equation}
 and
\begin{equation}
 t_{\rm spin-down}^{GR} \propto A_{\rm hb}^{-6/5} \tilde\Omega_f^{-8/5}.
\end{equation}
Assuming the final point is on the r-mode stability curve
\begin{equation}
\tilde \Omega_f^4 T_9^{-2} A_{\rm hb} = \tilde \Omega_f^6
\end{equation}
Using this expression together with Eq.\  (\ref{scaleTAhb}) we obtain $\Omega_f\propto A_{\rm hb}^{3/16}$ and $t_{\rm spin-down}^{GR} \propto A_{\rm hb}^{-3/2}$.

Typically, hyperon bulk viscosity is larger than boundary layer viscosity for the inertial modes. However, for the r-mode
boundary layer viscosity can be important at low $\tilde \Omega$.  For a $T_h = 2 \times 10^9$ K evolution without boundary layer 
viscosity with reduction factors included,  we obtain a scaling of $t_{\rm spin-down}^{GR} \propto A_{\rm hb}^{-1.3}$. 
When boundary layer viscosity is included,  $\Omega_f$ becomes fairly independent of $A_{\rm hb}$ changing only by about 
1\% when $A_{hb}$ is lowered by a factor of two. In this case the spin-down time approximately scales as
 \begin{equation}
t_{\rm spin-down}^{GR} \propto A_{\rm hb}^{-1}.
\end{equation}

In type III evolutions, when we assume that $1/\tan \phi^2 >> 1$, 
 \begin{eqnarray}
t_{\rm spin-down}^{GR} &&\propto \int_{\tilde\Omega_i = 0.67}^{\tilde\Omega_f} \int d\tilde\Omega \tilde\Omega^{-7} \\ \nonumber
&& \propto \tilde \Omega_f^{-6}.
 \end{eqnarray}
 On the $C=H$ curve
 \begin{equation}
\tilde \gamma_{GR} \tilde \Omega^2 \propto T_9^6
 \end{equation}
 So, $T_9 \propto \tilde \Omega^{4/3}$. On the r-mode stability curve, assuming boundary
 layer viscosity dominates
 \begin{equation}
 S_{\rm ns}^2 \frac{\sqrt{\tilde \Omega}}{T_9} \propto \tilde \Omega^6.
 \end{equation}
 The spin-down time scales as
 \begin{equation}
 t_{\rm spin-down}^{GR} \propto S_{\rm ns}^{-72/41}.
 \end{equation}
 When we include the reduction factors in the $T_h = 1.2 \times 10^{10}$ case we obtain a
similar scaling of approximately $ t_{\rm spin-down}^{GR} \propto S_{\rm ns}^{-2}$.
%At the intersection of the r-mode stability curve and the $C=H$ curve
% \begin{equation}
% \tilde\Omega \propto A_{\rm bl}^{6/31}
% \end{equation}
% and $t_{\rm spin-down}^{GR} \propto A_{\rm bl}^{-36/31}$.
 
To approximate the total timescale one can use a simple interpolation
\begin{equation}
t_{\rm spin-down} \approx[1/t_{\rm spin-down}^{GR}+1/t_{\rm spin-down}^{MD}]^{-1},
\end{equation}
where
\begin{eqnarray}
t_{\rm spin-down}^{MD} = 3.35 \times 10^8 B_{12}^{-2} (\Omega_f^{-2} - \Omega_i^{-2}) \; \rm sec.
\end{eqnarray}
The magnetic dipole spin-down timescale is approximately independent of $A_{\rm hb}$ for type I evolutions. 
%\begin{equation}
%\frac{d \tilde\Omega}{dt} = -\frac{1}{2} K_{GR} \tilde\Omega^5 - \frac{1}{2} K_{MD} \tilde\Omega^3, 
%\end{equation}
%where we assume $K_{GR}$ and $K_{MD}$ are constant with respect to $\tilde \Omega$. This approximation is correct for type I evolutions
%because $\gamma_\beta$ and $\gamma_\gamma$ are independent of $\tilde \Omega$.

% Making the change
%of variables  $x = \tilde \Omega^{-2}$ and solving for $t(x)$
%\begin{equation}
%t(x) = \frac{1}{K_{MD}} [x_f - x(0)] - \frac{K_{GR}}{K_{MD}^2} \log \left(\frac{K_{MD} x_f + K_{GR}}{K_{MD} x(0) + K_{GR}} \right)
%\end{equation}

%Assuming $x(0)$ to be small and defining
%\begin{eqnarray}
%t_{GR} = \left(\frac{1}{2} K_{GR} \tilde \Omega^4\right)^{-1}, \\ \nonumber
%t_{MD} =  \left(\frac{1}{2} K_{MD} \tilde \Omega^4\right)^{-1},
%\end{eqnarray}
%we obtain
%\begin{eqnarray}
%t_{\rm spin-down} = \frac{1}{2} t_{MD} \left[ 1 - \frac{t_{MD}}{t_{GR}} \log \left(1 + \frac{t_{GR}}{t_{MD}}\right) \right].
%\end{eqnarray}
%When gravitational radiation dominates 
%\begin{equation}
%t_{\rm spin-down} = \frac{1}{4} t_{GR},
%\end{equation}
%and when magnetic dipole radiation dominates
%\begin{equation}
%t_{\rm spin-down} = \frac{1}{2} t_{MD}.
%\end{equation}

\end{document}